\documentclass[twocolumn,secnumarabic,amssymb, nobibnotes, superscriptaddress, nofootinbib,aps,prd]{revtex4-2}
\usepackage[colorlinks,citecolor=blue,linkcolor=blue,anchorcolor=blue,filecolor=blue, urlcolor=blue]{hyperref}
\usepackage{natbib}

\usepackage{amsmath}
\usepackage{adjustbox}
\usepackage{graphicx}
\usepackage{booktabs}       
\usepackage{amsfonts}      
\usepackage{nicefrac}       
\usepackage{microtype}
\usepackage{cleveref}
\usepackage{xcolor}
\usepackage{multirow}
\usepackage{makecell}
\usepackage{array}
\usepackage{orcidlink}
\usepackage{float}
\usepackage{lineno}
\usepackage{slashed}
\usepackage{bm}
\definecolor{dora}{RGB}{255,150,0}

\usepackage[normalem]{ulem}
\usepackage{caption}
\DeclareCaptionJustification{fulljustify}{\leftskip=0pt \rightskip=0pt \parfillskip=0pt plus 1fil}
\usepackage{tikz}
\usepackage{tikz-feynman}
\usepackage{caption}
\usepackage{subcaption}

\begin{document}
\author{Adamu Issifu \orcidlink{0000-0002-2843-835X}} 
\email{ai@academico.ufpb.br}
\affiliation{CFisUC, Department of Physics, University of Coimbra, 3004-516 Coimbra, Portugal}
\affiliation{Departamento de F\'isica, Instituto Tecnol\'ogico de Aeron\'autica, DCTA, 12228-900, S\~ao Jos\'e dos Campos, SP, Brazil} 
\affiliation{Laborat\'orio de Computa\c c\~ao Cient\'ifica Avan\c cada e Modelamento (Lab-CCAM), Brazil}

\author{Constan\c ca Provid\^encia \orcidlink{0000-0001-6464-8023}}
\email{cp@uc.pt}
\affiliation{CFisUC, Department of Physics, University of Coimbra, 3004-516 Coimbra, Portugal}

\author{Franciele M. da Silva \orcidlink{0000-0003-2568-2901}} 
\email{franciele.m.s@ufsc.br}
\affiliation{Departamento de F\'isica, CFM - Universidade Federal de Santa Catarina, \\ Caixa Postal 5064, CEP 880.35-972, Florian\'opolis, SC, Brazil.}
\affiliation{Theoretical Astrophysics, Institute for Astronomy and Astrophysics, University of T\"{u}bingen, 72076 T\"{u}bingen, Germany}

\author{D\'ebora P. Menezes \orcidlink{0000-0003-0730-6689}}
\email{debora.p.m@ufsc.br}
\affiliation{Departamento de F\'isica, CFM - Universidade Federal de Santa Catarina, \\ Caixa Postal 5064, CEP 880.35-972, Florian\'opolis, SC, Brazil.}

\author{Tobias Frederico \orcidlink{0000-0002-5497-5490}} 
\email{tobias@ita.br}
\affiliation{Departamento de F\'isica, Instituto Tecnol\'ogico de Aeron\'autica, DCTA, 12228-900, S\~ao Jos\'e dos Campos, SP, Brazil}
\affiliation{Laborat\'orio de Computa\c c\~ao Cient\'ifica Avan\c cada e Modelamento (Lab-CCAM), Brazil}

\title{A self-consistent single-fluid framework for neutron stars admixed with mirror dark matter}

\begin{abstract} 
  We develop a self-consistent dark matter admix neutron star framework based on a contact vector current–current interaction that couples the chemical potentials of both sectors through mutual mean-field shifts, with the dark matter (DM) fraction $F_D = N_D/N_B$ fixed as a global input parameter. This formulation provides a physically motivated alternative to fixed-density prescriptions, allowing the local DM density to follow the baryonic matter (BM) density throughout the stellar interior. As an application, we consider a mirror-DM scenario with exact symmetry between the dark and visible sectors and investigate NS matter using the NL3$\omega\rho$, FSU2R,  NL3, and DDME2 equations of state (EOSs). We find that the amount of DM introduced through $F_D$ weakens the binding of dense matter, reduces its incompressibility, and softens the EOS, while the DM--BM interaction governs the microscopic behavior of the DM in the dense BM medium. Consequently, DM increases the central density and compactness of NSs, lowers their maximum masses, and shifts the onset of the direct Urca process to higher stellar densities. As a consequence, the onset of rapid cooling is shifted to more massive stars for models with a stiff symmetry energy and to less massive stars for models with a soft symmetry energy, depending on the extra compactness that results from the DM admixture. These results demonstrate that mirror-DM admixtures modify both the microscopic composition and macroscopic structure of NSs, with potential implications for their thermal evolution and multimessenger observational signatures.
\end{abstract}

\maketitle

\section{Introduction}
According to the standard cosmological model, the Universe is composed of roughly $5\%$ ordinary (baryonic) matter (BM), $\sim 27\%$ dark matter (DM), and $\sim 68\%$ dark energy. Strong evidence for DM arises from galactic rotation curves~\cite{1980ApJ...238..471R, 1981AJ.....86.1825B}, gravitational lensing~\cite{Clowe:2006eq}, galaxy cluster dynamics~\cite{Clowe:2006eq}, large-scale structure formation~\cite{Springel:2005nw, Planck:2015fie}, cosmic microwave background anisotropies~\cite{Planck:2015fie}, and observations of colliding galaxy clusters such as the Bullet Cluster. Totani~\cite{Totani:2025fxx} identified a $20\,\mathrm{GeV}$ gamma-ray excess in the Milky Way halo consistent with a Navarro–Frenk–White-like DM annihilation signal from Weakly Interacting Massive Particles (WIMPs) with masses $m_\chi \simeq 0.5\text{--}0.8\,\mathrm{TeV}$. Dark baryon black holes have also been introduced in \cite{Profumo:2025var}. However, despite the overwhelming astrophysical evidence for DM, its direct detection in ground-based experiments remains elusive~\cite{XENON:2018voc, XENON:2025vwd, LZ:2024zvo, PandaX:2024qfu}. Hence, the properties of DM, such as its mass, coupling strength, and mode of interaction, remain open research questions. 

The elusive nature of DM has motivated investigations of its effects in compact objects such as NSs and white dwarfs~\cite{Kouvaris:2007ay, Issifu:2024htq, Lopes:2024ixl, deLavallaz:2010wp}. Their extreme densities and strong gravitational fields make them promising astrophysical laboratories for indirect DM detection. Moreover, compact stars formed in high-DM-density environments, such as dwarf galaxies, galactic centers, and galaxy clusters, may accumulate significant amounts of DM over time and provide stronger observational signatures than terrestrial experiments~\cite{Kouvaris:2010vv, Bertone:2004pz}.

The limited knowledge of DM properties introduces uncertainties in how it should be modeled in compact objects. Consequently, two main approaches are commonly employed, one motivated by cosmology and the other by particle physics. In the cosmological picture, DM interacts with BM only through gravity. This is described within the two-fluid formalism, where the DM and BM EOSs are computed separately, while their interaction is introduced gravitationally through the two-fluid Tolman–Oppenheimer–Volkoff (TOV) equations~\cite{Shakeri:2022dwg, Das:2020ecp, Issifu:2025jac, Thakur:2023aqm, Karkevandi:2021ygv, Issifu:2025gsq, Kain:2021hpk, Biesdorf:2024dor, Rezaei:2016zje}. This framework allows for scenarios in which DM forms either a compact DM core inside the star or an extended DM halo surrounding it. In the particle-physics approach, DM interacts nongravitationally with BM particles through scattering processes. As DM particles traverse the star, repeated scatterings reduce their kinetic energy below the stellar escape velocity, allowing them to become gravitationally trapped and accumulate inside the compact object~\cite{Bertoni:2013bsa, Kouvaris:2007ay, Bramante:2013nma}. This is modeled through the single-fluid approach, where DM is coupled to the BM through a mediating particle or a contact interaction~\cite{Das:2018frc, Issifu:2025qqw, Das:2020vng, Gresham:2018rqo, Sahoo:2025rqw}.

The focus of the present work is to develop a self-consistent single-fluid framework that controls the DM content in DM-admixed neutron stars (DANSs) at realistic levels~\cite{Garani:2018kkd, Bramante:2023djs}. Existing single-fluid approaches typically estimate the DM abundance by fixing the DM Fermi momentum, $k_F^\chi$, as an external parameter. Since the corresponding number density, $n_\chi={(k_F^\chi)^3}/{3\pi^2}$, remains nearly uniform throughout the star, the relative DM fraction, $n_\chi/n_B$, decreases toward the high-density stellar core (see, e.g., Fig.~3 of~\cite{Sen:2021wev}). This prescription has been adopted in studies of feebly interacting DM~\cite{Sen:2021wev,Guha:2024pnn}, Higgs-portal DM~\cite{Panotopoulos:2017idn,Das:2018frc}, self-interacting fermionic DM~\cite{Narain:2006kx}, and relativistic mean-field descriptions of DANSs~\cite{Das:2020vng,Lopes:2024ixl}. However, such a density profile is difficult to reconcile with realistic capture scenarios, in which scattering, thermalization, and gravitational accumulation are expected to couple the DM content to the local baryonic environment and the capture history of the star~\cite{Kouvaris:2007ay, Bertoni:2013bsa, Bramante:2013nma}. Another adopted approach imposes neutron-decay equilibrium through the condition $\mu_\chi=\mu_n$~\cite{Motta:2018bil, Baym:2018ljz, Husain:2022bxl}. While useful as a limiting scenario, this prescription can generate DM populations comparable to the neutron population without explicitly accounting for finite capture rates, scattering cross sections, thermalization processes, or gravitational settling timescales~\cite{Bell:2023ysh, Berryman:2023rmh, Garani:2020wge}, potentially leading to an overestimation of the DM abundance inside NSs if a strong repulsive self-interaction is not considered.

To address these limitations, we consider a mirror-DM framework in which the dark sector is assumed to be an exact copy of the baryonic sector, with identical particle content and interactions, as motivated by the mirror-matter scenario of \cite{Oliveira:2015tha}. Both sectors are described within the same RMF formalism and are coupled through a vector current--current interaction that generates mutual shifts in their effective chemical potentials. In contrast to the WIMP-capture scenario of \cite{Kouvaris:2007ay}, where DM affects NSs only through capture and annihilation heating, the mirror DM framework treats DM as a dynamically coupled component interacting with BM, thereby directly modifying the EOS, particle composition, and global stellar properties. The framework minimizes the arbitrariness associated with the choice of coupling parameters and also relates the DM--BM interaction strength, $\alpha$, directly to the spin-independent DM--nucleon scattering cross section, $\sigma_{DN}$, constrained by terrestrial direct-detection experiments~\cite{XENON:2023cxc, XENON:2025vwd, CRESST:2019jnq, CRESST:2017cdd, PandaX:2024qfu, LZ:2024zvo, SuperCDMS:2018gro, SuperCDMS:2017mbc, SuperCDMS:2017nns}. This unique feature provides a direct bridge between compact star (CS) observables and particle-physics constraints, allowing NS measurements to be interpreted within the same parameter space probed by direct-detection experiments. 

The coupled equations are solved self-consistently by fixing the DM particle fraction, $F_D={N_D}/{N_B}$, where $N_D$ and $N_B$ denote the total DM and baryonic particle numbers, respectively. Since both sectors occupy the same stellar volume, this prescription ensures that the local DM density scales with the baryon density throughout the star, thereby maintaining a controlled and physically consistent DM admixture. The resulting framework avoids the artificial density profiles associated with fixed-$k_F^\chi$ models and the potentially excessive DM populations predicted by the neutron-decay-equilibrium scenario, if a strong repulsive self-interaction is not introduced, while providing a thermodynamically consistent description of the coupled dark and baryonic sectors.

The paper is organized as follows. In \cref{model}, we present the theoretical framework. The mirror-DM EOS is discussed in \cref{dmeos}, while the BM EOS is described in \cref{bmeos}. The properties of stellar matter incorporating different EOSs are presented in \cref{star}. In \cref{dmom}, we examine the effective saturation properties of DM-admixed symmetric nuclear matter and investigate how the presence of DM modifies the bulk properties of dense matter. Our results are presented in \cref{results}: the microscopic properties are discussed in \cref{micro}, the macroscopic stellar properties in \cref{macro}, and the interplay between the DM and BM sectors in \cref{dmomint}. Finally, the main conclusions, their implications for astrophysical observations and DM searches, and future perspectives are summarized in \cref{conclusions}.

\section{The Model} \label{model}

\begin{table}[b]
\centering
\caption{RMF model parameters used in this work. Masses are given in MeV.}
\begin{tabular}{c c c c}
\hline
Parameter & NL3$\omega\rho$~\cite{Providencia:2012rx} & FSU2R~\cite{Tolos:2016hhl, Tolos:2017lgv} & NL3~\cite{Horowitz:2000xj} \\
\hline
$m_N$      & 938.0   & 939.0 & 938.0 \\
$m_\sigma$ & 508.194 & 497.479 & 508.194  \\
$m_\omega$ & 782.501 & 782.500 & 782.501 \\
$m_\rho$   & 763.0   & 763.0 & 763.0 \\
$g_\sigma$ & 10.217  & 10.372 & 10.217 \\
$g_\omega$ & 12.868  & 13.505 & 12.868\\
$g_\rho$   & 11.277  & 14.367& 8.948 ($\sqrt{94.30}$) \\
$b$        & 0.00205 & 0.00165 & 0.00205\\
$c$        & -0.00265 & -0.00028 & -0.00265 \\
$\xi$      & 0.0     & 0.024  & 0.0 \\
$\Lambda_\omega$ & 0.03 & 0.045 & 0.0 (0.0125)\\
$n_0$ (fm$^{-3}$) & 0.148 & 0.1505 &0.148\\
\hline
\end{tabular}
\label{tabnl3}
\end{table}

\begin{table}[b]
\caption {DDME2 parameters for $m_N = 939.0$ MeV~\cite{PhysRevC.71.024312}.}
\begin{center}
\begin{tabular}{ |c| c| c| c| c| c| c| }
\hline
 meson($i$) & $m_i(\text{MeV})$ & $a_i$ & $b_i$ & $c_i$ & $d_i$ & $g_{i N} (n_0)$\\
 \hline
 $\sigma$ & 550.1238 & 1.3881 & 1.0943 & 1.7057 & 0.4421 & 10.5396 \\  
 $\omega$ & 783 & 1.3892 & 0.9240 & 1.4062 & 0.4775 & 13.0189  \\
 $\rho$ & 763 & 0.5647 & $\cdots$ & $\cdots$ & $\cdots$ & 7.3672 \\
 \hline
\end{tabular}
\label{tabddme2}
\end{center}
\end{table}

We develop a mirror DM model in which the dark sector interacts directly with the BM sector through a contact current--current interaction. Within this framework, the dark sector exactly mirrors the BM sector in particle masses, composition, coupling constants, and interaction channels. We employ three different EOSs within the relativistic mean-field (RMF) framework: the density-dependent meson-exchange model DDME2~\cite{PhysRevC.71.024312}, the Florida State University model version 2 with reduced neutron-star radii (FSU2R)~\cite{Tolos:2017lgv}, the third version of a NonLinear RMF (NL3)~\cite{Horowitz:2000xj}, and the nonlinear RMF model with an $\omega$--$\rho$ coupling, NL3$\omega\rho$~\cite{Providencia:2012rx, Horowitz:2000xj}. The model parameters used for the FSU2R, NL3, and NL3$\omega\rho$ models are given in \cref{tabnl3}, and the DDME2 model parameters are presented in \cref{tabddme2}. Explicitly, the models governing the study are given by equations (\ref{dme}) to (\ref{bd}):
\begin{widetext}
DM Lagrangian density:
\begin{align}\label{dme}
\mathcal{L}_D &= \bar{\psi}_D \Big[i\gamma^\mu \partial_\mu - g_{vD}\gamma^0 V_0- g_{\tilde{\rho} D}\gamma^0 \tilde{I}_{3\tilde{\rho}} \tilde{\rho}_{03}- \left(m_D - g_{sD}\tilde{\sigma}\right)
\Big]\psi_D - \frac{1}{2} m_{\tilde{\sigma}}^2 \tilde{\sigma}^2 + \frac{1}{2} m_v^2 V_0^2 + \frac{1}{2} m_{\tilde{\rho}}^2 \tilde{\rho}_{03}^{\,2} \nonumber\\ 
&- \frac{1}{3} \tilde{b}\, m_D \, g_{sD}^3 \tilde{\sigma}^3- \frac{1}{4} \tilde{c}\, g_{sD}^4 \tilde{\sigma}^4 + \frac{\tilde{\xi}}{4!} g_{vD}^4V_0^4+ \tilde{\Lambda}_\omega g_{\tilde{\rho}}^2 g_{vD}^2 \tilde{\rho}_{03}^2V_0^2, \nonumber\\
\mathcal{L}_{l_D} &=\bar{\psi}_{l_D}\left(i\gamma^\mu\partial_\mu -m_{l_D}\right)\psi_{l_D}.
\end{align}
The BM Lagrangian density:
\begin{align} \label{lgbm}
\mathcal{L}_B &=  \bar{\psi}_B \left[i\gamma^\mu\partial_\mu-\gamma^0 \left(  g_\omega \omega_0 + g_\rho I_{3\rho} \rho_{03} \right)- \left(m_N - g_\sigma \sigma_0 \right)\right] \psi_B - \frac{1}{2} m_\sigma^2 \sigma^2_0 + \frac{1}{2} m_\omega^2 \omega_0^2 + \frac{1}{2} m_\rho^2 \rho_{03}^2 \nonumber\\&
- \frac{1}{3} b\, m_N \, g_\sigma^3 \sigma^3_0- \frac{1}{4} c\, g_\sigma^4 \sigma^4_0+ \frac{\xi}{4!} g_\omega^4\omega_0^4+ \Lambda_\omega g_\rho^2 g_\omega^2 \rho_{03}^2\omega_0^2, \nonumber\\
\mathcal{L}_{l} &=\bar{\psi}_{l}\left(i\gamma^\mu\partial_\mu -m_{l}\right)\psi_{l}.
\end{align}
\end{widetext}
The interaction between DM and the BM:
\begin{equation}\label{bd}
    \mathcal{L}_{\rm DB} = \alpha \left(\bar{\psi}_D \gamma^\mu \psi_D\right)\left(\bar{\psi}_B \gamma_\mu \psi_B\right).
\end{equation}

The field $\psi_B(\bar{\psi}_B)$ is a Dirac spinor describing the nucleon doublet (proton and neutron) with nucleon mass $m_N$; $\gamma^\mu$ are the Dirac matrices, and $I_{3\rho}=\pm 1/2$ is the isospin projection. The couplings $g_\sigma$, $g_\omega$, and $g_\rho$ correspond to the meson fields $\sigma$, $\omega$, and $\rho$, with masses $m_\sigma$, $m_\omega$, and $m_\rho$, respectively. The nonlinear parameters $b$, $c$, $\xi$, and $\Lambda_\omega$, together with the couplings $g_i$, determine the bulk properties of the EOS. In particular, $b$ and $c$ regulate the incompressibility at saturation, increasing values of $\xi$ soften the EOS at high density, and $\Lambda_\omega$ reduces the slope of the symmetry energy at saturation~\cite{Providencia:2012rx,  Tolos:2017lgv}. 

For the density dependent model DDME2, the couplings $g_\sigma$, $g_\omega$, and $g_\rho$ are a function of the baryon density,
\begin{equation}
g_i(\rho) = g_{i,0}\,h_i(x), \quad x = n_B/n_0, \quad i = \sigma, \omega, \rho,
\label{eq:dd_coupling}
\end{equation}
with $g_{i,0}$ the couplings at saturation density $n_0$, and
\begin{equation}
h_i(x) = a_i\,\frac{1 + b_i(x + d_i)^2}{1 + c_i(x + d_i)^2}, \quad i=\sigma, \, \omega.
\label{eq:ddh_h}
\end{equation}
\begin{equation}
h_\rho(x) = \exp[-a_\rho(x-1)] ,
\label{eq:hy}
\end{equation}
and the parameters $b$, $c$, $\xi$, and $\Lambda_\omega$ are zero. For the DM, we consider the same density dependence of the couplings on the DM density. The parameters of the model are given in Table \ref{tabddme2}.

The dark-sector Lagrangian in \cref{dme} is obtained from \cref{lgbm} by introducing a tilde over the corresponding fields and nonlinear terms while preserving the same parameter values. Accordingly, the couplings $g_\sigma$, $g_\omega$, and $g_\rho$ in the BM sector are identified with $g_{sD}$, $g_{vD}$, and $g_{\tilde{\rho}D}$ in the dark sector, respectively. The total Lagrangian density of the system then becomes:
\begin{equation}
    \mathcal{L} = \mathcal{L}_{\rm D} + \mathcal{L}_{\rm B} + \mathcal{L}_{\rm DB} + \mathcal{L}_{\rm l_D} + \mathcal{L}_{\rm l},
\end{equation}
where $\mathcal{L}_{\rm D}$ and $\mathcal{L}_{\rm l_D}$ describe the dark baryons and dark leptons, with $l_D=(e',\, \mu')$ the dark leptons, respectively, while $\mathcal{L}_{\rm B}$ and $\mathcal{L}_{\rm l}$ correspond to the visible baryons and leptons. The term $\mathcal{L}_{\rm DB}$ represents the interaction between the dark and visible sectors.

The derivation of the EOS for the visible-matter sector in nonlinear RMF models has been presented in detail in Refs.~\cite{Providencia:2012rx, Tolos:2017lgv}; for a comprehensive review, see Ref.~\cite{Menezes:2021jmw}. The derivation of the EOS in density-dependent models differs slightly due to the presence of the rearrangement self-energy term, which is required to ensure thermodynamic consistency. Detailed discussions of this formalism can be found in Refs.~\cite{Typel:1999yq, Fuchs:1995as, Sedrakian:2022kgj}. In the present work, we do not reproduce these standard derivations. Instead, we focus on the EOS of the dark sector and the corresponding modifications induced in the BM sector in the following sections.

\subsection{Effective field theory interpretation}
Instead of introducing the BM--DM interaction phenomenologically through a contact current--current operator, let us assume that it is mediated by a generic massive vector boson $X_\mu$. The corresponding Lagrangian density \cite{Abercrombie:2015wmb, Langacker:2008yv, Abdallah:2015ter} is
\begin{align}
\mathcal{L}_{X}=&-\frac14 X_{\mu\nu}X^{\mu\nu}+\frac12 m_X^2X_\mu X^\mu - g_{NX}\bar{\psi}_B\gamma^\mu\psi_BX_\mu
\nonumber\\
&-g_{DX}\bar{\psi}_D\gamma^\mu\psi_DX_\mu,
\label{Vector}
\end{align}
where $X_{\mu\nu}=\partial_\mu X_\nu-\partial_\nu X_\mu$, $m_X$ is the mediator mass, and $g_{NX}$ and $g_{DX}$ denote its couplings to baryons and DM, respectively. In the low-energy regime relevant for NSs ($q^2\ll m_X^2$, with $q^2$ momentum transfer inside the star), the equation of motion reduces to
\begin{equation}
\label{Veq}
X^\mu\simeq
\frac{g_{NX}J_B^\mu+g_{DX}J_D^\mu}{m_X^2},
\end{equation}
where
$J_B^\mu=\bar{\psi}_B\gamma^\mu\psi_B$ and $J_D^\mu=\bar{\psi}_D\gamma^\mu\psi_D$ are the baryonic and DM currents. Substituting \cref{Veq} into \cref{Vector} and retaining only the mixed BM--DM interaction yields
\begin{equation}
\label{Xl}
\mathcal{L}_{\rm int}=-\frac{g_{NX}g_{DX}}{m_X^2}(\bar{\psi}_D\gamma^\mu\psi_D)(\bar{\psi}_B\gamma_\mu\psi_B).
\end{equation}

Although the interaction in \cref{bd} was introduced phenomenologically, it admits a simple ultraviolet completion. If the BM--DM interaction is mediated by a heavy vector boson $X_\mu$, integrating out the mediator in the low-energy limit $q^2\ll m_X^2$ generates the effective current--current interaction of \cref{Xl}. The sign of the effective coupling, $g_{DX}$, is determined by the relative $U(1)_X$ charge assignments of the baryonic and dark sectors. Choosing the dark fermion to carry the opposite $U(1)_X$ charge, corresponding to $g_{DX}<0$, gives
\begin{equation}\label{eq:alpha}
    \alpha=\frac{g_{NX}g_{DX}}{m_X^2},
\end{equation}
which reproduces the attractive BM--DM interaction adopted in \cref{bd}. Hence, the contact interaction employed in the present work may be interpreted as the low-energy effective-field-theory limit of a broad class of ultraviolet-complete vector-mediated DM models, including theories with heavy $Z'$ bosons \cite{Langacker:2008yv}, dark photons \cite{Holdom:1985ag, Essig:2013lka, Oliveira:2015tha}, and collider-inspired simplified DM models \cite{Abdallah:2015ter}. This correspondence provides a direct mapping between the effective coupling $\alpha$ entering the NS EOS and the microscopic parameters $(g_{NX},g_{DX},m_X)$, thereby establishing a common framework for connecting collider searches, underground direct-detection experiments, and NS observations. 

We note that, while our motivation stems from the mirror model of Ref.~\cite{Oliveira:2015tha}, formulated at the quark and lepton levels within an SU(3) Yang-Mills framework, we use a single U(1) gauge boson to mediate interactions between the visible and dark nucleons~\cite{Abercrombie:2015wmb, Langacker:2008yv, Abdallah:2015ter}. Furthermore, in a mirror model, we should adopt that $g_{NX}=g_{DX}$. However, this assumption is irrelevant to the current study because the key variable in our effective field theory approach is $\alpha$ in \cref{eq:alpha}.

\subsection{Dark matter equation of state} \label{dmeos}
The equation of motion obtained by varying the Lagrangian, \cref{dme} and \cref{bd}, with respect to $\bar{\psi}_D$ is
\begin{equation}
    \left[ i \gamma^\mu \partial_\mu - \gamma^0 \left( g_{VD} V_0 + g_{\tilde{\rho}} I_{3{\tilde{\rho}}} {\tilde{\rho}}_{03} - \alpha n_B \right) -  m_D^* \right] \psi_D = 0,
\end{equation}
where
\begin{equation}
    m_D^* = m_D - g_{sD}\tilde{\sigma}_0, \qquad n_B = \langle \bar{\psi}_B \gamma^0 \psi_B \rangle.
\end{equation}
The interaction between the BM and the DM sectors modifies the kinetic sector by $\alpha n_B$, which introduces a shift in the corresponding chemical potential
\begin{equation}
\mu_{D_i}=\sqrt{k_{Fi}^{2}+m_D^{*2}}+ g_{vD}V_0+ g_{\tilde{\rho}D}\,\tilde{I}_{3\tilde{\rho}}\,\tilde{\rho}_{03}- \alpha n_B.
\end{equation}
This yields the effective chemical potential 
\begin{equation*}
    \mu_{D_i}^*=\mu_{D_i} - g_{vD} V_0  - g_{\tilde{\rho} D}\tilde{I}_{3\tilde{\rho}} \tilde{\rho}_{03} + \alpha n_B,
\end{equation*}
where $\tilde{I}_{3n'}=-1/2$ and $\tilde{I}_{3p'}=1/2$. The field equations of the scalar mesons are:
\begin{align}
m_{\tilde{\sigma}}^2 \tilde{\sigma}+\tilde{b}\,m_D\,g_{sD}^{3}\tilde{\sigma}^{2}+\tilde{c}\,g_{sD}^{4}\tilde{\sigma}^{3}&=g_{sD}\,n_{sD},
\\
m_{v}^{2} V_{0}+\frac{\tilde{\xi}}{6}g_{vD}^{4}V_{0}^{3}+2\tilde{\Lambda}_{\omega}g_{vD}^{2}g_{\tilde{\rho}D}^{2}V_{0}\tilde{\rho}_{03}^{\,2}&=g_{vD}\,n_{D},
\\
m_{\tilde{\rho}}^{2}\tilde{\rho}_{03}+2\tilde{\Lambda}_{\omega}g_{\tilde{\rho}D}^{2}g_{vD}^{2}\tilde{\rho}_{03}V_{0}^{2}&=g_{\tilde{\rho}D}\,n_{3D},
\end{align}
where 
\begin{align}
n_D &= \sum_{i\in D}\frac{\gamma_i}{6\pi^2}k_{Fi}^{3},
\\
n_{sD} &= \sum_{i\in D}\frac{\gamma_i}{2\pi^2}\int_{0}^{k_{Fi}}\frac{m_D^{*}}{\sqrt{k^2+m_D^{*2}}}\,k^{2}\,dk,
\\
n_{3D} &= \sum_{i\in D}\tilde{I}_{3\tilde{\rho}}\dfrac{\gamma_i}{6\pi^2}k_{Fi}^{3},
\end{align}
with $i = n', p'$ representing dark neutrons and dark protons, respectively. Here, $n_D$, $n_{sD}$, and $n_{3D}$ denote the dark vector, scalar, and isovector densities, respectively. The summation runs over all dark baryon species in the mirror sector. The quantity $\gamma_i$ is the spin degeneracy factor of species $i$, $k_{Fi}$ is its Fermi momentum, and $\tilde{I}_{3\tilde{\rho}}$ is the third component of its isospin. The quantity $m_D^*$ represents the effective dark-baryon mass in the medium. 
\begin{widetext}
The dark-sector EOS is then obtained from
\begin{align}
\varepsilon_D &= \sum_{i={n}',{p}'}\frac{1}{\pi^2}\int_0^{k_{Fi}} dk \, k^2 \sqrt{k^2+m_D^{*2}}+\sum_{l_D}\frac{1}{\pi^2}\int_0^{k_{F,l_D}} dk \, k^2 \sqrt{k^2+m_{l_D}^{2}}
+\frac{1}{2}m_{\tilde{\sigma}}^2\tilde{\sigma}^2+\frac{1}{2}m_v^2V_0^2 \nonumber\\&+\frac{1}{2}m_{\tilde{\rho}}^2\tilde{\rho}_{03}^2+\frac{1}{3}\tilde{b}\,m_D(g_{sD}\tilde{\sigma})^3+\frac{1}{4}\tilde{c}(g_{sD}\tilde{\sigma})^4
\quad+\frac{\tilde{\xi}}{8}g_{vD}^4V_0^4+3\tilde{\Lambda}_\omega g_{\tilde{\rho}D}^{\,2}g_{vD}^{\,2}\tilde{\rho}_{03}^{\,2}V_0^2,
\\
P_D &= \sum_{i={n}',{p}'}\frac{1}{3\pi^2}\int_0^{k_{Fi}} dk \,\frac{k^4}{\sqrt{k^2+m_D^{*2}}}+\sum_{l_D}\frac{1}{3\pi^2}\int_0^{k_{F,l_D}} dk \,
\frac{k^4}{\sqrt{k^2+m_{l_D}^{2}}}
-\frac{1}{2}m_{\tilde{\sigma}}^2\tilde{\sigma}^2+\frac{1}{2}m_v^2V_0^2+\frac{1}{2}m_{\tilde{\rho}}^2\tilde{\rho}_{03}^2\nonumber\\&-\frac{1}{3}\tilde{b}\,m_D(g_{sD}\tilde{\sigma})^3
-\frac{1}{4}\tilde{c}(g_{sD}\tilde{\sigma})^4\quad+\frac{\tilde{\xi}}{{24}}g_{vD}^4V_0^4+\tilde{\Lambda}_\omega g_{\tilde{\rho}D}^{\,2}g_{vD}^{\,2}\tilde{\rho}_{03}^{\,2}V_0^2,
\end{align}
where $\varepsilon_D$ is the energy density of the dark sector and $P_D$ is its corresponding pressure, excluding the DM--BM interaction.
\end{widetext}

For NS matter, dark leptons are treated as free Fermi gases with chemical potentials
\begin{equation}
\mu_{{l_D}} = \sqrt{k_{F,l_D}^2 + m_{{l_D}}^2}, \qquad l_D = e', \mu'.
\end{equation}
Dark $\beta$ equilibrium is established through the weak process
\begin{equation}
n' \rightarrow p' + e' + \bar{\nu}'.
\end{equation}
The corresponding chemical equilibrium condition is
\begin{equation}
\mu_{{n'}} = \mu_{{p'}} + \mu_{{e'}} - \mu_{{\nu'}}.
\end{equation}
Assuming that dark neutrinos escape freely ($\mu_{{\nu'}} = 0$), we obtain
\begin{equation}
\mu_{{n'}} = \mu_{{p'}} + \mu_{{e'}}.
\end{equation}
Additionally, the equilibrium between dark leptons requires
\begin{equation}
\mu_{{e'}} = \mu_{{\mu'}}.
\end{equation}
The dark sector satisfies electric charge neutrality:
\begin{equation}
n_{{p'}} = n_{{e'}} + n_{{\mu'}}.
\end{equation}

The $F_D$ is defined as the global ratio of total particle numbers~\cite{Hajkarim:2024ecp, 1996csnp.book.....G},
\begin{equation}\label{dmf}
    F_D = \frac{N_D}{N_B},
\end{equation}
with 
\begin{align}
    N_D =& \int_0^R 4\pi r^2\left(1-\frac{2M(r)}{r}\right)^{-1/2}n_D(r)dr \quad{\rm and}\quad\\ N_B =& \int_0^R 4\pi r^2\left(1-\frac{2M(r)}{r}\right)^{-1/2}n_B(r)dr \label{NbNd}
\end{align}
where $r$ is the radial coordinate measured from the center of the star, $R$ is the stellar radius, and $M$ is its total mass, including the DM and BM mass. Here, the proper volume element is
\begin{equation}
dV_p=
4\pi r^2
\left(1-\frac{2M(r)}{r}\right)^{-1/2}dr.
\end{equation}
Within the present single-fluid framework, we assume that the captured DM has thermalized with the BM and co-moves with the stellar fluid. We interpret this as an effective description of efficient DM capture, thermalization, and subsequent mixing with the baryonic fluid, thereby redistributing the captured DM so that it remains co-moving with the BM. Additionally, we impose a constant-composition condition, namely, the local DM-to-BM particle ratio is fixed in every co-moving fluid element,
\begin{equation}
\frac{dN_D}{dN_B}=F_D,
\end{equation}
with
\begin{equation}
dN_D=n_D(r)dV_p, \qquad dN_B=n_B(r)dV_p.
\end{equation}
Consequently,
\begin{equation}
\frac{n_D(r)dV_p}{n_B(r)dV_p}=F_D,
\end{equation}
which gives
\begin{equation}
\label{fd}
n_D(r)=F_D n_B(r).
\end{equation}

Thus, $F_D$ specifies the total DM content of the star while, under the constant-composition assumption, it also fixes the local DM-to-BM density ratio. The local $n_D(r)$ and corresponding $k_F^\chi$ are
therefore determined from the $n_B(r)$ profile through \cref{fd}, rather than by independently prescribing $k_F^\chi$. This prescription provides the closure relation for the present single-fluid model. It ensures that the DM follows the spatial distribution of the BM, becoming more concentrated toward the stellar core where $n_B(r)$ is largest.

The local densities $n_D$ and $n_B$ are obtained self-consistently from the coupled mean-field equations, where the contact interaction $\alpha(\bar{\psi}_D\gamma^\mu\psi_D)(\bar{\psi}_B\gamma_\mu\psi_B)$ modifies the chemical potentials of both sectors through mean-field shifts $-\alpha n_D$ in the baryonic sector and $-\alpha n_B$ in the dark sector, dynamically coupling the two fluids. The system is solved simultaneously with baryon number conservation and $\beta$-equilibrium imposed separately on each sector. The resulting EOS, constructed from the coupled BM and DM components under the constraint $n_D = F_D n_B$, is then used to solve the TOV equations and determine the structure of the DANSs.

\subsection{Baryonic matter equation of state} \label{bmeos}

In the BM sector, variation of the Lagrangian density, \cref{lgbm}, with respect to the adjoint nucleon field $\bar{\psi}_B$ yields the Dirac-like equation
\begin{equation}\label{eqB}
    \left[ i\gamma^\mu\partial_\mu- \gamma^0\left(g_\omega\omega_0+ g_\rho I_{3\rho}\rho_{03}- \alpha n_D\right) - m_N^* \right]\psi_B = 0,
\end{equation}
where $I_{3\rho}=+1/2$ for protons and $I_{3\rho}=-1/2$ for neutrons. The nucleon effective mass is generated by the scalar meson field and is given by
\begin{equation}
    m_N^* = m_N - g_\sigma \sigma.
\end{equation}
In the mean-field approximation, the vector meson fields and the DM background modify the single-particle energies through shifts in the vector potential. Thus, the nucleon chemical potentials obtained from the Fermi-surface energies are given by:
\begin{equation}
    \mu_{n,p} =\sqrt{k_{F,n,p}^{2}+m_N^{*2}} + g_\omega\omega_0+ g_\rho I_{3B}\rho_{03}- \alpha n_D,
    \label{mub}
\end{equation}
where $k_{F,n,p}$ denotes the neutron and proton Fermi momenta, respectively, and $n_D$ is the DM number density. Notice that for DDME2, the chemical potential \cref{mub} contains an extra rearrangement term \cite{Typel:1999yq}. The term proportional to $\alpha$ arises from the DM--BM interaction and introduces an additional shift in the BM chemical potentials. It is convenient to define the effective chemical potentials,
\begin{equation}
    \mu_{n,p}^{*}=\mu_{n,p}- g_\omega\omega_0- g_\rho I_{3\rho}\rho_{03}+ \alpha n_D,
\end{equation}
such that
\begin{equation}
    \mu_{n,p}^{*}=\sqrt{k_{F,n,p}^{2}+m_N^{*2}}.
\end{equation}
The baryon number densities are then given by
\begin{equation}
    n_{n,p}=\frac{k_{F,n,p}^{3}}{3\pi^{2}},
\end{equation}
and the total baryon density is
\begin{equation}
    n_B = n_n + n_p.
\end{equation}

\subsection{Stellar matter properties} \label{star}
The interaction term \cref{bd}
introduces a physically important modification to the chemical potentials of both sectors. In the mean-field approximation, $\mathcal{L}_{DB}^{MF}=\alpha n_Bn_D$, the interaction contributes with terms proportional to the density of the opposite fluid, leading to the shifts
\begin{equation}
\mu_B \rightarrow \mu_B - \alpha n_D,
\qquad\text{and}\qquad
\mu_D \rightarrow \mu_D - \alpha n_B.
\end{equation}
These terms represent the mean-field interaction energy generated by the DM and BM backgrounds, respectively. It quantifies the change in the energy required to add a particle to either sector in the presence of the other. Consequently, the BM and DM chemical equilibria become dynamically coupled, so that the local properties of each fluid depend on the density of the other. This mechanism enables a self-consistent determination of the DM and BM densities throughout the star while maintaining a prescribed global DM fraction. Unlike phenomenological approaches based on a fixed $k_F^\chi$, which artificially enforce an approximately uniform DM density, or neutron-decay equilibrium models that impose $\mu_D=\mu_n$, a mechanism that is independent of the capture history and dynamical response of the dark sector. The present framework allows both fluids to adjust their chemical potentials and densities self-consistently. As a result, the DM content remains under consistent thermodynamic equilibrium throughout the stellar interior.

Furthermore, the coupling strength $\alpha$ can be related directly to the dark-nucleon--visible-nucleon scattering cross section, $\sigma_{DN}$, through
\begin{equation}
\sigma_{DN}=\frac{4\alpha^2 m_D^2 m_N^2}{\pi (m_D+m_N)^2}=\frac{4\alpha^2 \mu_{DN}^2}{\pi},
\end{equation}
where
\begin{equation}
\mu_{DN}=\frac{m_D m_N}{m_D+m_N}
\end{equation}
is the dark-nucleon--visible-nucleon reduced mass. For the mirror DM model considered in this work, where $m_D=m_N$, the coupling constant can be expressed as
\begin{equation}
\alpha=\sqrt{\frac{\pi \sigma_{DN}}{m_D^2}}.
\end{equation}
This relation provides a direct connection between the stellar properties and particle-physics constraints. To constrain the interaction strength, we compare $\sigma_{DN}$ with the experimental limits reported by the CRESST-III collaboration~\cite{CRESST:2019jnq, CRESST:2017cdd}, which probes DM particles in the mass range $0.1$--$1.0$ GeV. From the reported exclusion curves, the upper limit corresponding to a DM mass of approximately $1$ GeV is $\sigma_{DN}\sim10^{-38}\,\mathrm{cm}^2$ (upper limits from other direct DM detection groups can be found in \cref{tab:dm_constraints}). This value is adopted as a benchmark constraint in our calculations to determine the corresponding coupling strength $\alpha$.

The CRESST-III upper limit on the spin-independent DM--nucleon scattering cross section is used to constrain the effective coupling $\alpha$ entering the NS EOS. We identify the relevant direct-detection component with the mirror nucleon $D$, taking $m_D= m_N$ in the exact mirror-symmetry limit. Thus, CRESST-III is used solely as an experimental upper bound on $\alpha$, rather than as a measured interaction strength.

Although $\alpha$ is small due to the weak BM--DM interaction strength, its value remains positive, i.e., $\alpha>0$. A positive $\alpha$ corresponds to an attractive, isospin-independent mean-field shift, $-\alpha n_D$, opposite in sign to the repulsive $\omega$-meson contribution, $+g_\omega\omega_0$. While the $\omega$ field increases the chemical potential required to introduce an additional baryon into dense matter, the BM--DM interaction lowers the chemical potentials of both the baryonic and mirror-dark sectors. As a result, the energy cost of adding DM to the baryonic sector and, by mirror symmetry, baryons to the dark sector is reduced, thereby favoring the coexistence of the two components. Although this common shift cancels in the $\beta$-equilibrium condition, it remains relevant through the charge-neutrality constraint and the self-consistent meson fields, driving the matter toward a more neutron-rich composition and ultimately softening the EOS. The full derivation of the dark-nucleon--visible-nucleon cross-section can be found in Appendix~\ref{apend}.

Therefore, after imposing both visible and dark-sector $\beta$-equilibria, together with electric charge neutrality in each sector, the total EOS is given by
\begin{align}
    P_{\rm tot} &= P_B + P_D + P_{BD},\qquad P_{BD} = \alpha n_B n_D,
    \\
    \varepsilon_{\rm tot} &= \varepsilon_B + \varepsilon_D + \varepsilon_{BD},\qquad \varepsilon_{BD} = -\alpha n_B n_D.
\end{align}
Here, $P_{\rm tot}$ and $\varepsilon_{\rm tot}$ denote the total pressure and energy density of the system, respectively. The quantities $P_B$ ($\varepsilon_B$) and $P_D$ ($\varepsilon_D$) are the pressures (energy densities) of the baryonic and dark sectors, respectively, while $P_{BD}$ and $\varepsilon_{BD}$ represent the contributions arising from the DM--BM interaction. The full derivation of $\varepsilon_{BD}$ and $P_{BD}$ are presented in Appendix~\ref{apdixB}. The opposite signs of the interaction terms in the pressure and energy density follow directly from the mean-field treatment of the vector current--current interaction.

A notable difference between the DDME2 parametrization and the nonlinear RMF models is the explicit density dependence of the meson--baryon couplings. As a result, variations of the couplings with density generate an additional rearrangement self-energy, which must be included to preserve thermodynamic consistency. In the dark sector, the rearrangement contribution is given by
\begin{equation}
\Sigma_R^D=
\frac{\partial g_{vD}}{\partial n_D}V_0\,n_D
+\frac{\partial g_{\tilde{\rho}D}}{\partial n_D}\tilde{\rho}_{03}\,n_{3D} - \frac{\partial g_{sD}}{\partial n_D}\tilde{\sigma}\,n_{sD},
\end{equation}
and enters the dark-baryon chemical potentials together with the vector mean fields and the BM--DM interaction term. An analogous rearrangement self-energy,
\begin{equation}
\Sigma_R^B=\frac{\partial g_{\omega}}{\partial n_B}\omega_0\,n_B
+\frac{\partial g_{\rho}}{\partial n_B}\rho_{03}\,n_{3B}- \frac{\partial g_{\sigma}}{\partial n_B}\sigma\,n_{sB},
\end{equation}
arises in the BM sector. Although the rearrangement term does not represent a new interaction, it is essential for thermodynamic consistency and the fulfillment of the Hugenholtz--Van Hove theorem~\cite{Hugenholtz1958Phy}. Therefore, whenever the DDME2 parametrization is employed, the rearrangement contributions are incorporated self-consistently in both the BM and DM sectors in the calculation of the EOS~\cite{Typel:1999yq, Fuchs:1995as, Sedrakian:2022kgj, Avancini:2003ur}.

\subsection{Effective saturation properties of dark matter--admixed symmetric nuclear matter} \label{dmom}

This section explores how the presence of a DM admixture modifies the effective bulk properties of symmetric matter. We emphasize that the quantities derived below should not be interpreted as the empirical saturation properties of ordinary nuclear matter, which correspond to purely BM. Rather, they characterize a hypothetical system composed of symmetric nuclear matter admixed with a finite DM fraction and provide insight into how DM may alter the bulk properties of dense matter.

For a fixed DM fraction $F_D$, the total particle density is given by
\begin{equation}
n_{\rm tot}=n_B+n_D=(1+F_D)n_B,
\end{equation}
where
\begin{equation}\label{nBnD}
n_D=F_D n_B.
\end{equation}
In the absence of DM ($F_D=0$), the standard saturation properties of symmetric nuclear matter associated with the underlying hadronic model are recovered. For $F_D>0$, the DM component contributes to the total energy density and pressure, thereby modifying the effective saturation properties of the combined system.

The energy per particle of the admixed matter is defined as
\begin{equation}
E_{\rm tot}(n_{\rm tot},\delta)
=\frac{\varepsilon(n_B,n_D,\delta_{B,D})}{n_{\rm tot}}-m_N,
\end{equation}
where $\delta_B=(n_n-n_p)/n_B$ is the isospin asymmetry parameter ($B$ and $D$ are used to differentiate between BM and DM isospin asymmetries), with $n_n$ and $n_p$, neutron and proton number densities, respectively. The effective saturation properties of symmetric matter ($\delta=0$) are then obtained from the behavior of $E_{\rm tot}(n_{\rm tot},0)$ as a function of the total density $n_{\rm tot}$. The saturation density $n_0$ satisfies
\begin{equation}
\left.\frac{dE_{\rm tot}}{dn_{\rm tot}}\right|_{n_{\rm tot}=n_0,\delta=0}=0,
\end{equation}
with the corresponding binding energy per particle
\begin{equation}
E_0=E_{\rm tot}(n_0,0),
\end{equation}
and incompressibility
\begin{equation}
K_0=9n_0^2\left.\frac{d^2E_{\rm tot}}{dn_{\rm tot}^2}\right|_{n_{\rm tot}=n_0}.
\end{equation}
These quantities provide a measure of how a DM admixture modifies the saturation behavior and stiffness of the effective symmetric nuclear matter.

The onset of the direct Urca (DU) process in $\beta$-equilibrated stellar matter is determined by the critical proton fraction
\begin{equation}
    x_{\rm DU}=\frac{1}{1+\left(1+x_e^{1/3}\right)^3},\qquad x_e =\frac{n_e}{n_e+n_\mu},
\end{equation}
where $n_e$ and $n_\mu$ denote the electron and muon number densities, respectively. The particle fractions are defined as
\begin{equation}
    Y_i=\frac{n_i}{n_B},
\end{equation}
where $i=(n,p,e,\mu)$ labels the particle species and $n_i$ is the corresponding number density. The DU process becomes kinematically allowed when the proton fraction exceeds the critical threshold,
\begin{equation}
    Y_p = \frac{n_p}{n_B}\geq x_{\rm DU}.
\end{equation}
The density at which this condition is first satisfied defines the onset of the direct Urca process in the stellar interior.

\section{Results} \label{results}
Having established the theoretical framework in the previous sections, we now turn to the numerical results. The amount of DM composition expected inside NSs is highly model dependent and remains largely unconstrained~\cite{Bramante:2023djs,Garani:2018kkd}, as it depends sensitively on the underlying DM particle properties and the mechanisms governing its production, capture, and accumulation in NSs, which remain poorly constrained by current observations and experiments. Although DM is estimated to constitute approximately $27\%$ of the total matter content of the Universe, the fraction that can accumulate inside NSs over their lifetime remains highly uncertain. Therefore, in this work, we treat $F_D$ as a free parameter and systematically explore values of $0\%$, $2\%$, $5\%$, and $10\%$ to quantify its impact on the EOS, symmetric nuclear matter properties, particle composition, and global NS observables.

\begin{table*}[t!]
\centering
\caption{Saturation properties of symmetric stellar matter in the presence of DM for different RMF models. Here, $F_D$ denotes the DM fraction; $n_0$ is the saturation density; $E_0$ is the total energy per nucleon; $K_0$ is the incompressibility. $m_N^*/m_N$ and $m_D^*/m_D$ are the nucleon and DM effective masses normalized to their bare masses, $\varepsilon_B$ and $\varepsilon_D$ are the BM and the DM energy densities of the effective symmetric nuclear matter admixed with DM.}
\begin{ruledtabular}
\begin{tabular}{c c c c c c c c c }
Model & $F_D\,$(\%) & $n_0\,$(fm$^{-3}$) & $E_0\,$(MeV) & $K_0\,$(MeV) & $m_N^*/m_N$ & $m_D^*/m_D$ & $\varepsilon_B\,(\rm MeV\,fm^{-3}$)  & $\varepsilon_D/\varepsilon_B$ \\
\hline

\multicolumn{9}{c}{\textbf{NL3$\omega\rho$}} \\
\hline
NL3$\omega\rho$ & 0  & 0.1477 & -16.178 & 270.83 & 0.5960 & $\cdots$& 136.1398 & 0.00 \\
    & 2  & 0.1477 & -15.862 & 265.65 & 0.5959 & 0.9902& 136.1767 &  0.0204 \\
    & 5  & 0.1481 & -15.451 & 259.36 & 0.5950 & 0.9756 & 136.4850 &  0.0508 \\
    & 10 & 0.1493 & -14.925 & 252.30 & 0.5921 & 0.9521 & 137.5949 &  0.1015 \\

\hline

\multicolumn{9}{c}{\textbf{FSU2R}} \\
\hline
FSU2R & 0  & 0.1504 & -16.265 & 237.69 & 0.5931 & $\cdots$& 138.7807 &  0.00 \\
       & 2  & 0.1504 & -15.946 & 233.08 & 0.5931 & 0.9893 & 138.8177&  0.0204  \\
       & 5  & 0.1509 & -15.538 & 227.59 & 0.5922 & 0.9737 & 139.2003 &  0.0508 \\
       & 10 & 0.1524 & -15.023 & 221.40 & 0.5891 & 0.9478 & 140.5957 & 0.1015 \\

\hline

\multicolumn{9}{c}{\textbf{DDME2}} \\
\hline
DDME2 & 0  & 0.1522 & -16.142 & 252.04 & 0.571 & $\cdots$& 140.4658 &  0.00 \\
      & 2  & 0.1522 & -15.827 & 247.05 & 0.571 & 0.986 & 140.4658 &  0.0203 \\
      & 5  & 0.1522 & -15.426 & 239.94 & 0.571 & 0.965 & 140.4658 &  0.0508 \\
      & 10 & 0.1539 & -14.948 & 236.39 & 0.568 & 0.931 & 142.0403 &  0.1014 \\

\end{tabular}
\end{ruledtabular}
\label{tabmicro}
\end{table*}

\begin{table*}[t!]
\scriptsize
\centering
\caption{ Macroscopic properties of DM-admixed NSs for  
NL3 with different symmetry energy slopes, including
NL3$\omega\rho$, FSU2R, and DDME2 EOSs at different DM fractions, $F_D$. The table shows the maximum gravitational mass, $M_{\rm max}$, the corresponding percentage mass variation relative to the $F_D=0$ configuration, $\Delta M_{\rm max}$, the radius of the maximum-mass configuration, $R$, the central stellar density, $n_{c,\rm max}$, and the central baryonic chemical potential, $\mu_c$. For canonical $1.4\,M_\odot$ stars, $R_{1.4}$ and $n_{c,1.4}$ denote the stellar radius and central stellar density respectively. The table also lists the direct Urca threshold stellar density, $n_{\rm DU}$, and the gravitational masses at the onset of the direct Urca process, $M_{\rm DU}$. {In addition to NL3$\omega\rho$, }{two {extra} NL3 parameterizations with different symmetry energy slopes were considered: the standard NL3 ($L=118$ MeV) and a modified version with a moderate slope ($L=82$ MeV),  with the results for the latter given in parentheses.} Increasing $F_D$ generally reduces the stellar radius and maximum mass while increasing the central density and impacting the stellar mass for the onset of direct Urca cooling {depending on the symmetry energy slope}.
}
\begin{ruledtabular}
\begin{tabular}{c c c c c c c c c c c}
Model & $F_D\,$(\%) & $M_{\rm max}\,$($M_\odot$) & $\Delta M_{\rm max}\,$(\%) & $R\,$(km) & $n_{c,\,\rm max}\,$(fm$^{-3}$) & $R_{1.4}\,$(km) & $n_{c,\,1.4}\,$(fm$^{-3}$) & $n_{\rm DU}\,$(fm$^{-3}$) & $M_{\rm DU}\,$($M_\odot$) & $\mu_c\,$(MeV)\\
\hline

\multicolumn{11}{c}{\textbf{DDME2} ($L=51$ MeV)} \\
\hline
DDME2 & 0  & 2.49 &  0.00 & 12.08 & 0.81 & 13.46 & 0.34 & $\cdots$ & $\cdots$ & 1901\\
       & 2  & 2.47 & -0.80 & 11.91 & 0.83 & 13.18 & 0.35 & $\cdots$ & $\cdots$ & 1939\\
       & 5  & 2.41 & -3.21 & 11.68 & 0.84 & 12.93 & 0.36 & $\cdots$ & $\cdots$ & 1956\\
       & 10 & 2.32 & -6.83 & 11.29 & 0.86 & 12.54 & 0.37 & $\cdots$ & $\cdots$ & 1991\\

\hline

\multicolumn{11}{c}{\textbf{FSU2R}($L=44$ MeV)} \\
\hline
FSU2R & 0  & 2.06 &  0.00 & 11.57 & 0.94 & 12.77 & 0.38 & 0.57 & 1.88 & 1587\\
       & 2  & 2.01 & -2.43 & 11.44 & 0.95 & 12.83 & 0.40 & 0.60 & 1.87 & 1591\\
       & 5  & 1.96 & -4.85 & 11.19 & 0.97 & 12.48 & 0.41 & 0.61 & 1.83  & 1604\\
       & 10 & 1.87 & -9.22 & 10.84 & 0.99 & 12.23 & 0.44 & 0.64 & 1.76 & 1620\\

\hline

\multicolumn{11}{c}{\textbf{NL3} ($L=118 (82)$ MeV)} \\
\hline
NL3 & 0  & 2.78(2.75) &  0.00 & 13.35(13.07) & 0.63(0.69) & 14.77(14.15) & 0.27 (0.29) & 0.20(0.29) & 0.80(1.40) & $\cdots$\\
       & 2  & 2.74(2.71) & -1.44(-1.45) & 13.16(12.90) & 0.65(0.69) & 14.53(13.93) & 0.28(0.30) & 0.21(0.30) & 0.83(1.47) & $\cdots$\\
       & 5  & 2.68(2.65) & -3.60(-3.64) & 12.88(12.64) & 0.67(0.71) & 14.22(13.67) & 0.28(0.30) & 0.21(0.31) & 0.83(1.47) & $\cdots$\\
       & 10 & 2.58(2.55) & -7.19(-7.27) & 12.48(12.23) & 0.71(0.72) & 13.76(13.34) & 0.29(0.31) & 0.22(0.32) & 0.85(1.53) & $\cdots$\\

\hline

\multicolumn{11}{c}{\textbf{NL3$\omega\rho$} ($L=55$ MeV)} \\
\hline
NL3$\omega\rho$
                & 0  & 2.76 &  0.00 & 13.03 & 0.68 & 14.05 & 0.29 & 0.51 & 2.65 & 1939\\
                & 2  & 2.72 & -1.45 & 12.85 & 0.69 & 13.85 & 0.30 & 0.52 & 2.61 & 1960\\
                & 5  & 2.66 & -3.62 & 12.59 & 0.70 & 13.60 & 0.30 & 0.54 & 2.58 & 1983\\
                & 10 & 2.56 & -7.25 & 12.20 & 0.72 & 13.21 & 0.31 & 0.56 & 2.49 & 2027\\

\end{tabular}
\end{ruledtabular}
\label{tabmacro}
\end{table*}

\Cref{tabmicro} and \cref{tabmacro} illustrate how a fixed $F_D$, modifies both the microscopic properties of dense matter and the macroscopic structure of NSs. At saturation density (see~\cref{tabmicro}), increasing $F_D$ produces a consistent trend across all RMF models: $n_0$ changes only marginally, while $E_0$ becomes progressively less negative, indicating a reduction in the overall binding of the system. Simultaneously, $K_0$ decreases by approximately $6{-}7\%$ between $F_D=0\%$ and $10\%$, demonstrating a systematic softening of the EOS. The nucleon effective mass ratio, $m_N^*/m_N$, remains nearly unchanged, whereas $m_D^*/m_D$ decreases more noticeably, reaching values of $0.93{-}0.95$ at $F_D=10\%$. In contrast, $\varepsilon_B$ increases only slightly, while $\varepsilon_D/\varepsilon_B$ grows in direct proportion to $F_D$, consistent with the imposed local constraint $n_D =F_Dn_B$. This implies that the dark sector contributes with an increasingly significant fraction of the $\varepsilon_{\rm tot}$. These results indicate that the BM--DM interaction weakens the binding of the system and reduces its resistance to compression without significantly affecting $n_0$.

The consequences of this softening are reflected directly in the stellar properties shown in \cref{tabmacro}. As $F_D$ increases from $0\%$ to $10\%$, $M_{\rm max}$ decreases by approximately $7\%$ for NL3$\omega\rho$ ($2.76 \rightarrow 2.56\,M_\odot$), $9\%$ for FSU2R ($2.06 \rightarrow 1.87\,M_\odot$), and $7\%$ for DDME2 ($2.49 \rightarrow 2.32\,M_\odot$). At the same time, both $R$ and $R_{1.4}$ decrease by roughly $0.8{-}1.0$ km, while $n_{c,\rm max}$ and $n_{c,1.4}$ increase, indicating that DANSs become more compact and centrally compressed. The central chemical potential, $\mu_c$, also increases monotonically with $F_D$, because a larger stellar central density ($n_c$) is needed to support the same or reduced mass, which is a direct consequence of EOS softening. Another important consequence is the modification of the DU process: for NL3$\omega\rho$ and FSU2R, $n_{\rm DU}$ shifts to larger values, implying that the onset of rapid neutrino cooling is delayed. In contrast, the DU process does not occur within the DDME2. Although the three models differ quantitatively, with NL3$\omega\rho$ producing the stiffest EOS and largest stellar masses and radii, DDME2 yielding intermediate predictions, and FSU2R generating the softest stars, they exhibit the same qualitative response to DM. Overall, the results show that a finite DM fraction softens the EOS, reduces the maximum supported mass, increases stellar compactness, and shifts the onset of rapid neutrino cooling to higher densities.

\subsection{Microscopic properties} \label{micro}

\begin{figure}[t!]
   \centering	
	\includegraphics[width=0.5\textwidth]{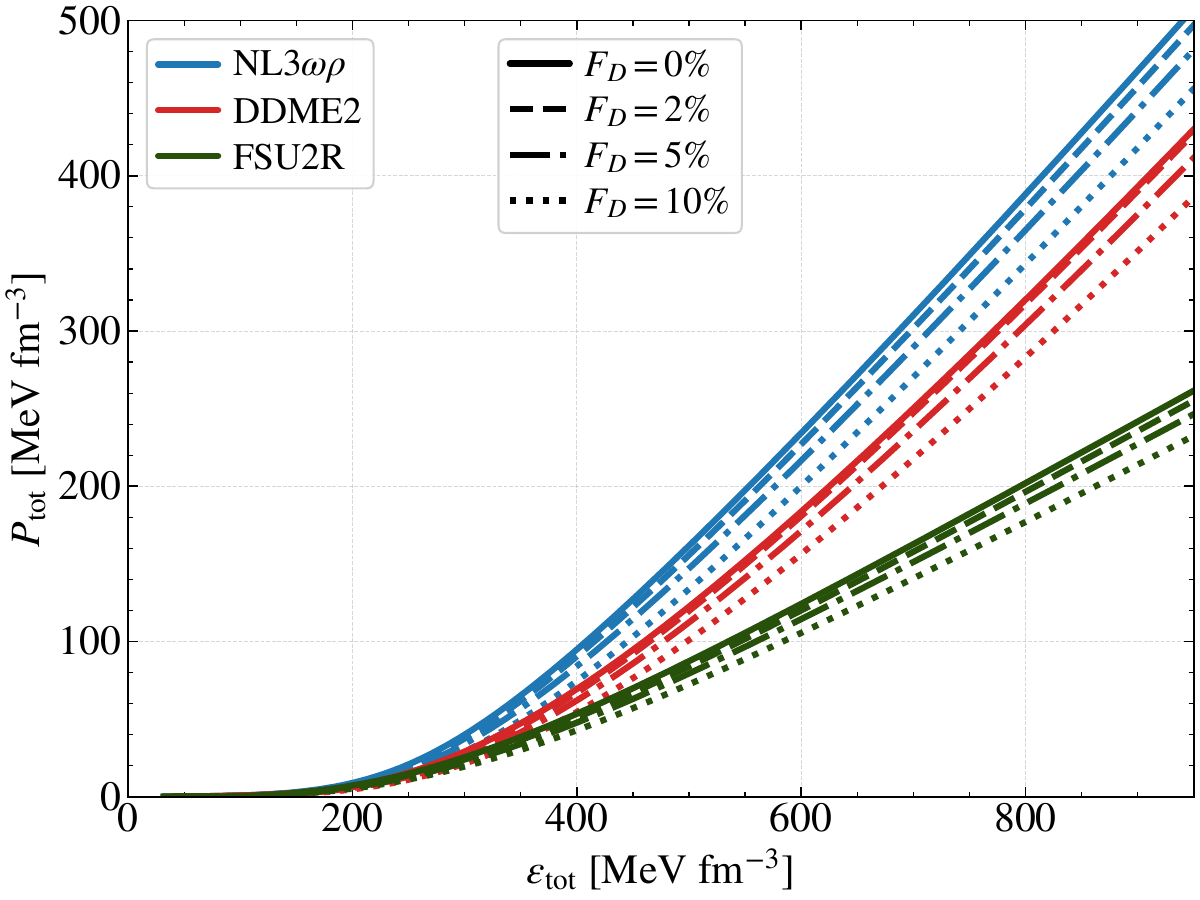}
    	\caption{Pressure as a function of the total energy density for DM-admixed matter described by the NL3$\omega\rho$, DDME2, and FSU2R EOSs. Colors denote the RMF parametrizations and line styles indicate the DM fraction $F_D$. The systematic downward shift of the curves with increasing $F_D$ demonstrates that the presence of DM softens the EOS. Among the models considered, NL3$\omega\rho$ predicts the highest pressures (stiffest EOS), FSU2R the lowest pressures (softest EOS), while DDME2 lies between these two limits.
        } 
\label{EOS_curve}
\end{figure}

The EOS curves in \cref{EOS_curve} show the dependence of $P_{\rm tot}$ on $\varepsilon_{\rm tot}$ for the NL3$\omega\rho$, DDME2, and FSU2R models at different $F_D$. A clear and systematic trend emerges: increasing $F_D$ shifts the EOS to lower $P_{\rm tot}$ at a fixed $\varepsilon_{\rm tot}$, indicating that the presence of DM softens the EOS. This behavior is consistent with the reduction in binding energy and incompressibility reported in \cref{tabmicro} and reflects the increasing contribution of the dark sector to the total energy density. The softening becomes more discernible at higher $\varepsilon_{\rm tot}$, where the separation between the curves grows, demonstrating that DM has its strongest impact in the dense cores of NSs. Consequently, DM-admixed matter offers less pressure support at a given energy density, leading to higher central densities and more compact stellar configurations. 

Despite this common DM-induced softening, the intrinsic stiffness hierarchy of the nuclear models is preserved. NL3$\omega\rho$ remains the stiffest EOS, producing the highest $P_{\rm tot}$ throughout the density range, DDME2 exhibits intermediate behavior, and FSU2R remains the softest. This is a consequence of considering mirror DM, which has the same properties as the BM sector. The stiffness hierarchy even becomes more clearly distinguishable at high density, which is visible in the figure. These differences directly translate into the stellar properties presented in \cref{tabmacro}, where NL3$\omega\rho$ supports the largest masses and radii, while FSU2R yields the most compact stars at a fixed mass (see the canonical $1.4\, M_\odot$ data for clarity). Overall, the figure demonstrates that the primary effect of DM is to soften the EOS irrespective of the underlying RMF model, the consequences of which are quantified in \cref{tabmacro} and \cref{tabmicro}.

    \begin{figure*}[t!]
   \centering	
	\includegraphics[width=\textwidth]{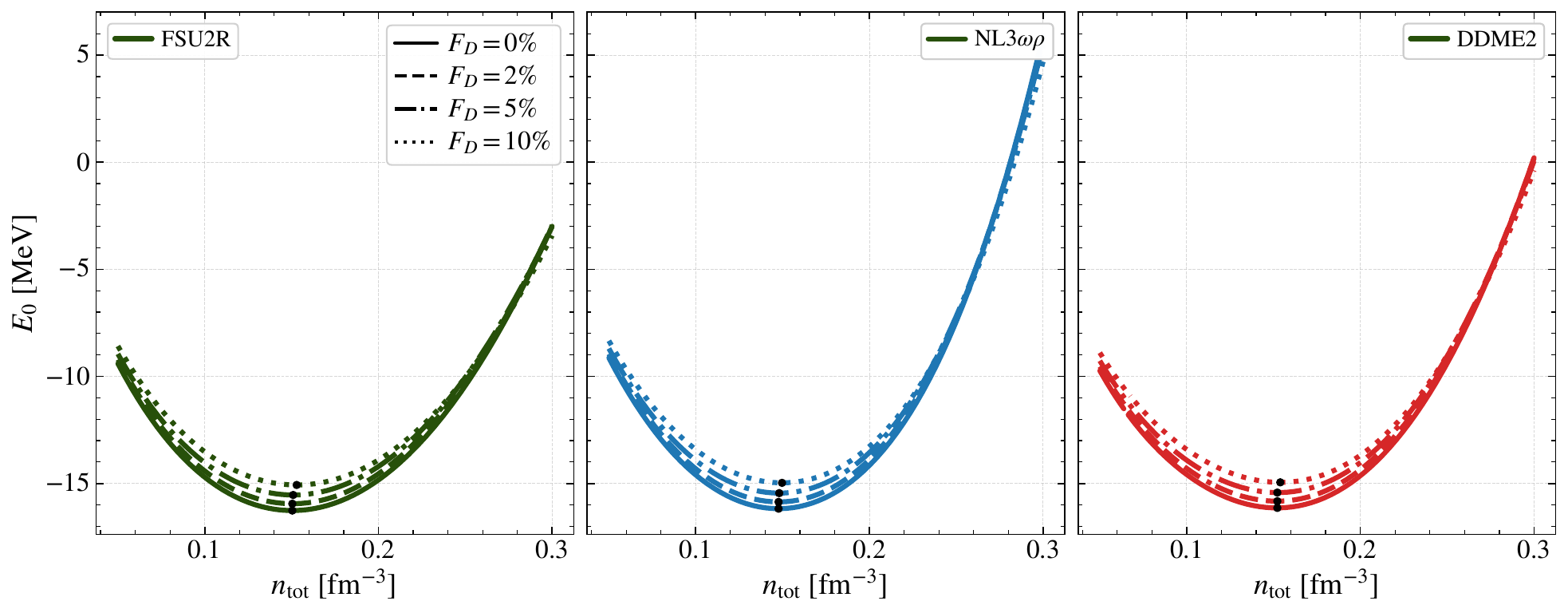}
     	\caption{Binding energy per nucleon as a function of total baryon density 
        for the NL3$\omega\rho$, FSU2R, and DDME2 models at different DM fractions $F_D$. Colors distinguish the RMF parametrizations, while line styles correspond to the adopted DM fractions. Increasing DM content makes the matter less bound and shifts the saturation point to slightly higher densities.} 
\label{SYM_curve_NL}
	\end{figure*}

\Cref{SYM_curve_NL} shows the binding energy per nucleon as a function of {$n_{\rm tot}$} for the NL3$\omega\rho$, FSU2R, and DDME2 models at different $F_D$. A common trend emerges across all three EOS parametrizations: increasing $F_D$ shifts the curves upward, making the $E_0$ less negative and therefore reducing the overall binding of the system. This behavior reflects the increasing contribution of the dark sector to {$n_{\rm tot}$, and $E_{\rm tot}$ that form the basics for determining $E_0$}. The saturation density $n_0$ is only weakly affected, increasing by less than $2\%$ up to $F_D=10\%$, consistent with the values reported in \cref{tabmicro}. The separation between the curves becomes slightly more pronounced above saturation density, indicating that the influence of DM grows in the high-density regime. Although the three models have nearly identical $E_0$ at $F_D=0\%$, with FSU2R exhibiting the deepest minimum, followed by NL3$\omega\rho$ and DDME2, they all display the same qualitative response to DM presence. The reduction in $E_0$ is accompanied by a systematic decrease in $K_0$ (as shown in \cref{tabmicro}), demonstrating that DM weakens the resistance of matter to compression. This modification of the bulk properties of dense matter provides the microscopic origin of the EOS softening and explains the reduction in maximum stellar mass and the increase in stellar compactness reported in \cref{tabmacro}. 

    \begin{figure*}[t!]
   \centering	
	\includegraphics[width=\textwidth]{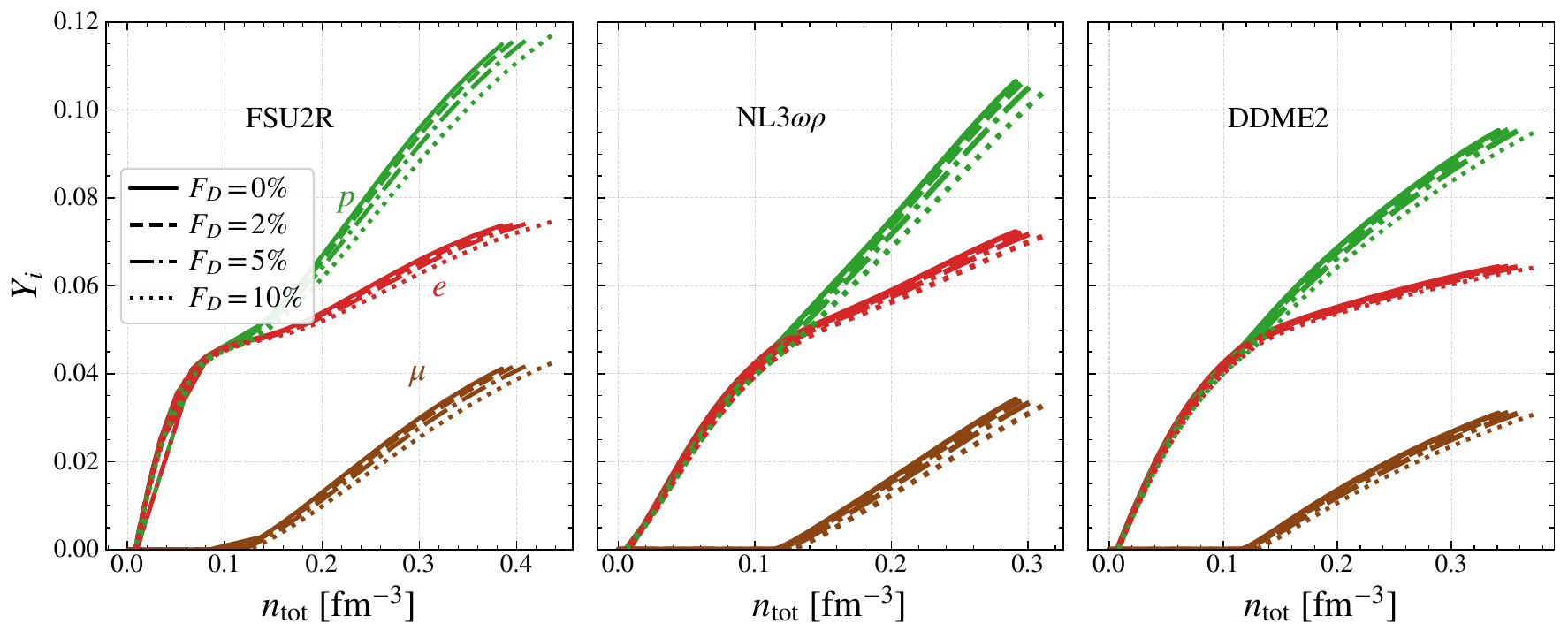}
    	\caption{Particle fractions as functions of total density, $n_{\rm tot}$, for canonical $1.4\,M_\odot$ stars described by the NL3$\omega\rho$, FSU2R, and DDME2 EOSs. Colors denote the particle species, while line styles indicate the DM fraction $F_D$. The presence of DM systematically modifies the composition of dense matter by changing the equilibrium abundances of protons and leptons.} 
\label{particle_f}
	\end{figure*}

\Cref{particle_f} shows the particle composition of a canonical $1.4\,M_\odot$ NS. The neutron fraction, $Y_n$, is not displayed because it constitutes more than $88\%$ across the density range $0.08 \lesssim n_{\rm tot} \lesssim 0.44~\mathrm{fm}^{-3}$ attained in the interior of a canonical $1.4\,M_\odot$ star. The proton, electron, and muon fractions reveal that increasing $F_D$ systematically suppresses the proton abundance at a given total density $n_{\rm tot}$. The composition is determined by the simultaneous requirements of $\beta$-equilibrium and charge neutrality in the presence of dark matter. Since only protons participate directly in the charge-neutrality condition through their coupling to the lepton sector, the system minimizes its free energy by reducing the proton and lepton abundances in favor of neutrons. Consequently, increasing the $F_D$ drives the matter toward a more neutron-rich composition, suppressing the proton and lepton fractions at a given stellar density. The effect becomes more pronounced toward the stellar core, where $n_{\rm tot}$ is highest and the influence of the BM--DM interaction is strongest. A lower $Y_p$  shifts the stellar onset density of nucleonic DU processes, $n_{\rm DU}$, to larger values. 

{The direct effect on the mass of the star with a central density equal to the DU onset density, $M_{\rm DU}$, depends on the density dependence of the symmetry energy: i) for a stiff symmetry energy (like NL3) the DU onset density is quite low and $M_{\rm DU}$ is shifted to larger masses, ii) for a soft symmetry energy (like NL3$\omega\rho$) the DU onset density is high and as a consequence $M_{\rm DU}$ is large or does not even occur (for DDME2 model considered). In this case, the extra compression of matter due to the presence of DM is large enough at these densities to shift the DU proton fraction to smaller star masses, i.e., smaller  $M_{\rm DU}$. This has been confirmed considering the NL3$\omega\rho$ model with different symmetry energy slopes. Although the EOS models differ quantitatively due to their distinct symmetry-energy properties, they exhibit the same qualitative response to DM. These results show that DM not only softens the EOS but also alters the composition of canonical NSs and modifies their thermal evolution.

\begin{figure}[h]
   \centering	
	\includegraphics[width=0.5\textwidth]{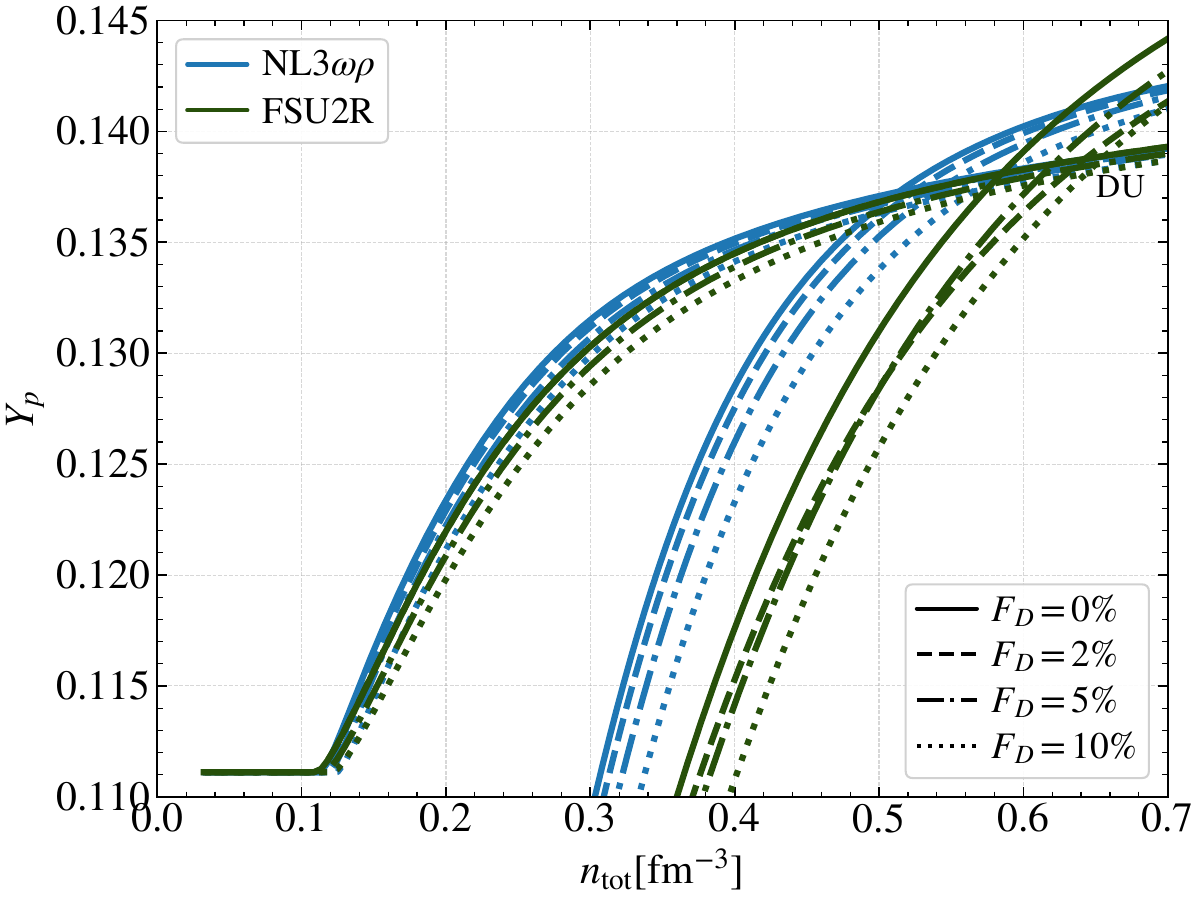}
    	\caption{Proton fraction and the corresponding direct Urca threshold as functions of total density ($n=n_D+n_B$) for the NL3$\omega\rho$ and FSU2R EOSs. Colors denote the RMF parametrizations, while line styles indicate the DM fractions $F_D$. Increasing DM content lowers the proton fraction and shifts the direct Urca onset to higher densities.
        } 
\label{DU_curve}
	\end{figure}

The $Y_p$ and DU-threshold curves in \cref{DU_curve} illustrate how DM modifies the composition and cooling properties of dense matter for two models that predict the opening of DU inside massive enough stars, FSU2R and NL3$\omega\rho$. $Y_p$ increases with density, as required by $\beta$-equilibrium. Increasing $F_D$ systematically suppresses $Y_p$ at a given density, shifting the $Y_p$ curves downward and moving their intersection with the DU threshold to progressively higher densities. This provides a direct microscopic explanation for the increase in $n_{\rm DU}$ for increasing $F_D$ reported in \cref{tabmacro}. Among the two models considered, FSU2R exhibits the largest $Y_p$ and therefore reaches the DU threshold at the lowest densities. In contrast, the $Y_p$ predicted by DDME2 remains below the DU threshold throughout the entire density range supported by its stellar configurations for all values of $F_D$ considered. Therefore, the DU process is never activated in this model, and consequently, no corresponding curves appear in  \cref{DU_curve}. The NL3 models included in \cref{tabmacro} show a similar qualitative behavior like the other models, except that the crossing of the DU proton fraction with the $\beta$-equilibrium proton fraction occurs at lower densities. These results are included as a qualitative consistency check to illustrate the connection between the microscopic properties of the EOS, especially the symmetry energy and its slope, and the onset of the DU process in NSs.

The NL3$\omega\rho$ model displays intermediate behavior, with $M_{\rm DU}$ ranging from $2.49\,M_\odot$ ($F_D=10\%$)  to $2.65\,M_\odot$ (no DM), implying that rapid cooling is restricted to higher densities and more compact stellar configurations. These trends are consistent with the broader NS literature~\cite{Lattimer:1991ib}, where mechanisms that reduce the proton abundance, such as a softer symmetry energy or the appearance of additional degrees of freedom, tend to delay or suppress the nucleonic DU process~\cite{Fortin:2016hny, Klahn:2006ir}. In the present model, DM produces a similar effect by modifying the $\beta$-equilibrium and charge neutrality conditions, driving the matter toward a more neutron-rich composition at a given stellar density. The figure, therefore, highlights an important astrophysical consequence of DM admixture: beyond softening the EOS and increasing stellar compactness, DM can significantly alter neutrino-emission channels and shift rapid cooling to higher densities. Depending on the increase in compactness, although the DU is shifted to higher stellar densities, it may occur in smaller or larger mass stars, depending on the density dependence of the symmetry energy, when admixed with DM, providing a potentially observable imprint on the thermal evolution of NSs~\cite{Page:2004fy, Page:2005fq}. Quantifying this effect and disentangling it from other composition-dependent cooling mechanisms, however, requires dedicated cooling simulations.
    
\subsection{Macroscopic properties} \label{macro}

    \begin{figure}[t!]
   \centering	
	\includegraphics[width=0.5\textwidth]{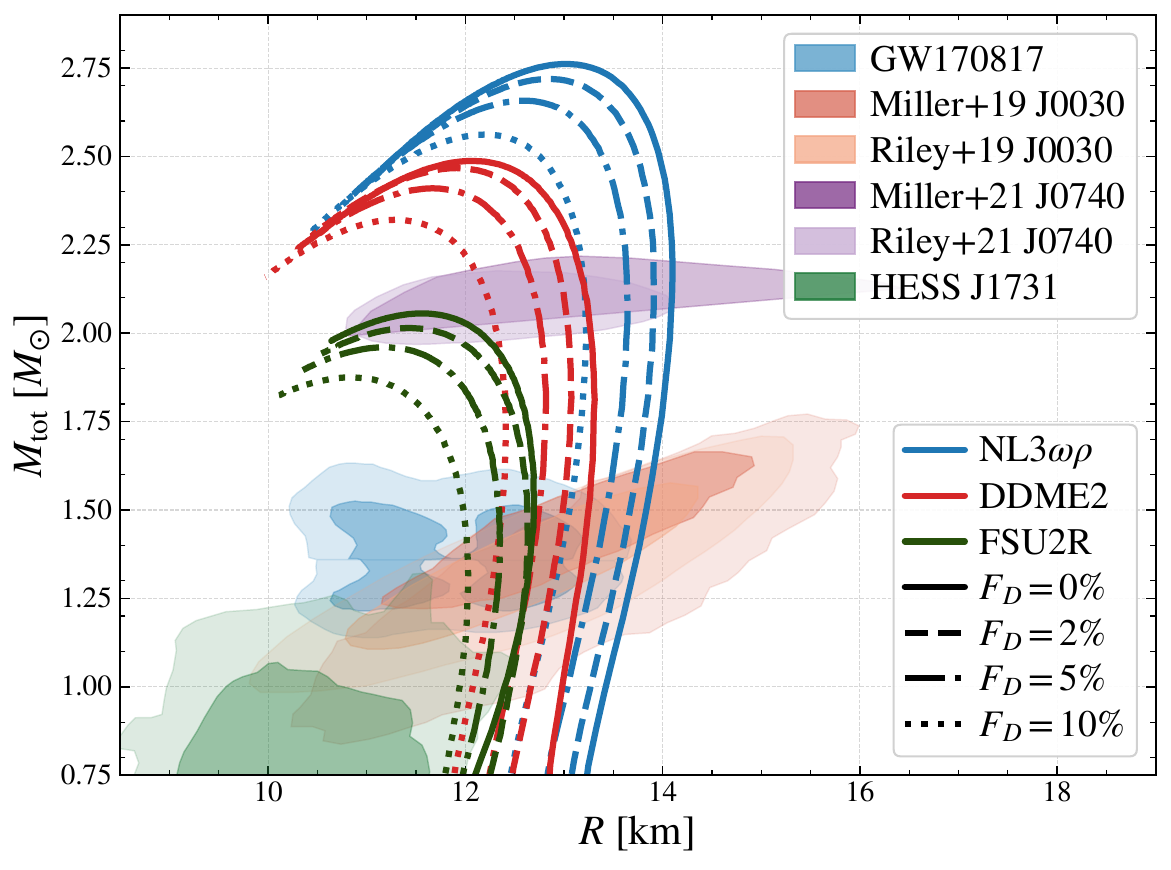}
    	\caption{The plot presents the total mass and radius relation for three nuclear matter EOSs with an admixture of DM. Colors denote the EOS models, while line styles indicate the corresponding DM particle fractions. Confidence contours show observational constraints from the secondary component of the binary merger event GW170817~\cite{abbott2019PhRvX} (blue; outer contour 90\% credible region (CR), inner contour 50\% CR), HESS~J1731--347~\cite{doroshenko2022strangely}(green; outer contour 95\% CR, inner contour 68\% CR), PSR~J0740+6620 (purple for Miller \textit{et al.}~\cite{miller2021} and lavender for Riley \textit{et al.}~\cite{riley2021}, both at 95\% CR), and PSR~J0030+0451 (brick for Miller \textit{et al.}~\cite{Miller:2019cac} and salmon for Riley \textit{et al.}~\cite{riley2019}, both at 95\% CR).} 
\label{MR_curve}
	\end{figure}

Figure \ref{MR_curve} shows the mass-radius relation for the three different EOS models studied in this work: NL3$\omega\rho$ (blue), DDME2 (red), and FSU2R (green), without DM (solid line) and with 2\% (dashed), 5\% (dot-dash), and 10\% (dotted) DM. In addition, the credibility regions for GW170817, HESS J1731$-$347, PSR J0740$+$6620, and PSR J0030$+$045 are also shown, as explained in the caption. As expected, we observe that the stiffest EOS (NL3$\omega\rho$) produces NSs with the highest maximum mass, while FSU2R, which is the softest EOS, produces stars with the lowest maximum mass, and DDME2 produces an intermediate maximum mass.
In the absence of DM, NL3$\omega\rho$ satisfies only the constraints from PSR J0740$+$6620 and PSR J0030$+$045, while DDME2 also satisfies the mass-radius region of GW170817, and FSU2R is the only EOS that satisfies all constraints in the absence of DM. The effect of DM is qualitatively the same for all EOSs considered: increasing $F_D$ softens the EOS and shifts the mass--radius curves toward lower masses and smaller radii.

As the $F_D$ increases, the BM--DM interaction contributes more significantly to the thermodynamics of dense matter, reducing the effective pressure support against gravitational collapse. As a result, the star must contract to a higher central density $n_c$ in order to maintain hydrostatic equilibrium, producing more compact configurations and reducing the maximum mass that can be supported. This softening of the EOS, together with the associated reductions in the maximum mass and stellar radius, is qualitatively consistent with the behavior reported in single-fluid~\cite{Das:2020vng,Gresham:2018rqo,Sahoo:2025rqw}, two-fluid~\cite{Thakur:2023aqm,Karkevandi:2021ygv,Issifu:2025gsq}, and neutron-decay DANS models~\cite{Baym:2018ljz,Husain:2022bxl}. Although the underlying coupling mechanisms differ among these scenarios, they generally lead to a softer EOS and more compact stellar configurations. These trends are fully consistent with the behavior of the stellar central density $n_c$ and radii reported in \cref{tabmacro}. 

The reduction in the maximum mass is most pronounced for FSU2R, the softest EOS considered, reaching approximately $9.2\%$ at $F_D=10\%$. In contrast, NL3$\omega\rho$ exhibits a smaller reduction of about $7.3\%$, while DDME2 is the least affected, with a decrease of only $6.8\%$. This hierarchy reflects the differing responses of the EOSs to DM-induced softening. In general, softer EOSs experience larger reductions in pressure support and therefore larger decreases in their maximum supported masses, whereas stiffer EOSs are more resistant to the effects of the BM--DM interaction. However, the maximum mass reduction can be model-dependent, since DDME2 is a softer EOS than NL3$\omega\rho$ but has a smaller reduction in $M_{\rm max}$ as the DM content increases. 

The reduction in radius is considerably less model dependent, with all three EOSs exhibiting a similar decrease of roughly $6\%$ for the maximum-mass configurations at $F_D=10\%$. As the $F_D$ increases, the NL3$\omega\rho$ mass--radius relation shifts toward the GW170817 probability distributions, although it remains incompatible with the HESS J1731$-$347 constraint even at the highest DM content considered. FSU2R moves deeper into the regions favored by both GW170817 and HESS J1731$-$347; however, the accompanying reduction in the maximum mass drives the model below the observational lower limit set by the $\sim 2\,M_\odot$ pulsars, particularly PSR J0740$+$6620. Among the EOSs studied, DDME2 provides the most favorable overall agreement with observations. As the $F_D$ increases, it remains consistent with the massive pulsar constraints while simultaneously moving closer to the probability distributions inferred from GW170817 and HESS J1731$-$347. This balance between maintaining a sufficiently large maximum mass and reproducing radius constraints highlights DDME2 as the EOS most compatible with the current multimessenger data in the presence of DM, {even though it requires a considerably large $F_D$ to satisfy the HESS J1731$-$347 CR}.

\begin{figure}[t!]
   \centering	
	\includegraphics[width=0.5\textwidth]{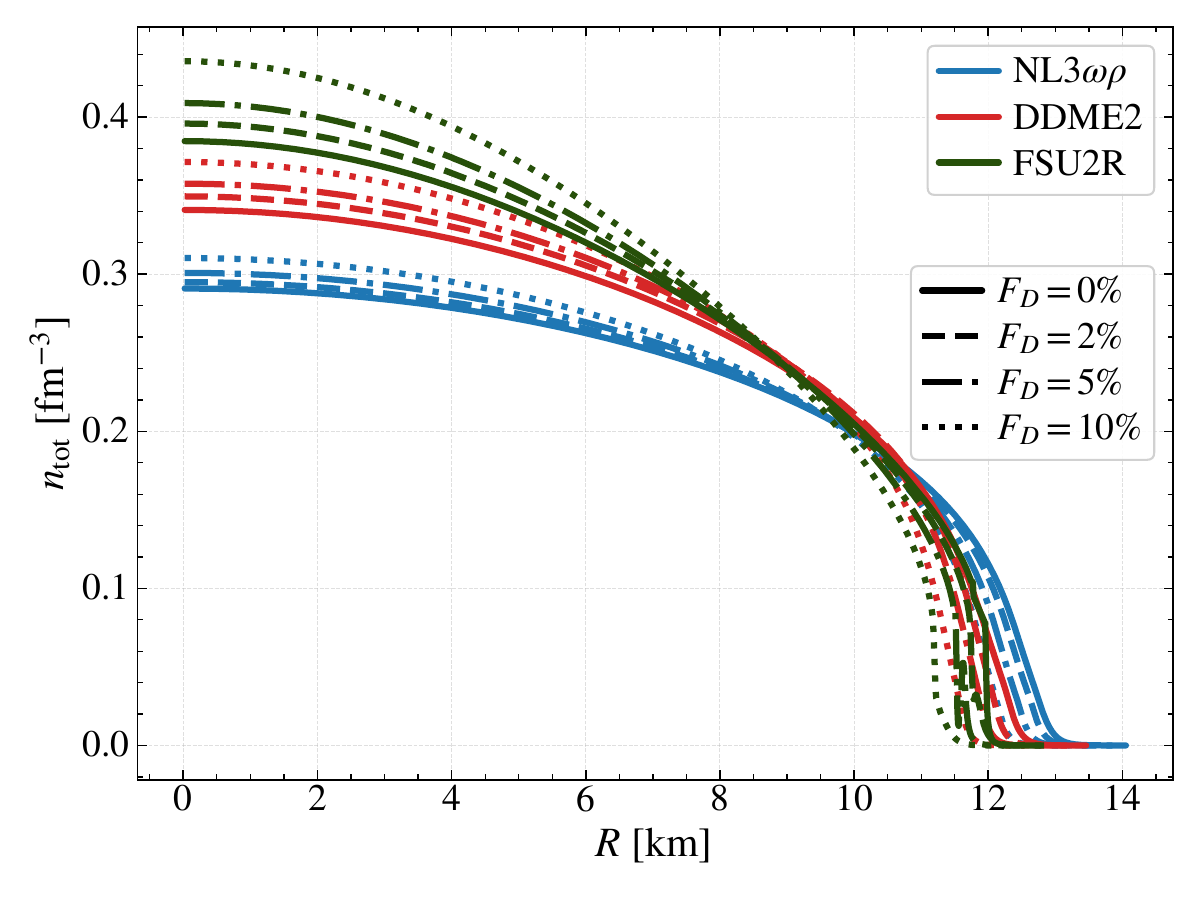}
    	\caption{{The total stellar  density} profiles of canonical $1.4\,M_\odot$ DANSs for the NL3$\omega\rho$, FSU2R, and DDME2 EOSs. Colors represent different BM models, and line styles correspond to the DM fractions $F_D$. The presence of DM increases the central density and reduces the stellar radius, illustrating the enhanced compactness induced by the DM admixture.
        } 
\label{nBxR}
	\end{figure}

The density profiles in \cref{nBxR} show the total density $n=n_{\rm tot}$ as a function of $R$ for a canonical $1.4\,M_\odot$ star described by the NL3$\omega\rho$, FSU2R, and DDME2 EOSs at different $F_D$. Across all models, increasing $F_D$ raises the total central density, $n_c$, and reduces the stellar radius, indicating progressively more compact configurations. Quantitatively, $n_c$ increases by approximately $7\%-16\%$, while the radius decreases by about $0.5$--$1$ km as $F_D$ rises from $0\%$ to $10\%$, consistent with the canonical star properties reported in \cref{tabmacro}. In addition, the density profiles become steeper as $F_D$ increases, reflecting enhanced central compression and reduced pressure support throughout the stellar interior. The hierarchy observed in the EOSs and stellar properties is clearly reproduced: NL3$\omega\rho$ yields the largest radii and lowest $n_c$, FSU2R produces the most compact stars with the highest $n_c$, and DDME2 exhibits intermediate behavior. These trends reflect the DM-induced softening of the EOS, whereby a higher $n_c$ is required to support the same gravitational mass. The figure, therefore, provides a direct spatial visualization of how DM modifies the internal structure and compactness of NSs.

\subsection{Interaction between dark and baryonic matter} \label{dmomint}

   \begin{figure}[t!]
   \centering	
	\includegraphics[width=0.5\textwidth]{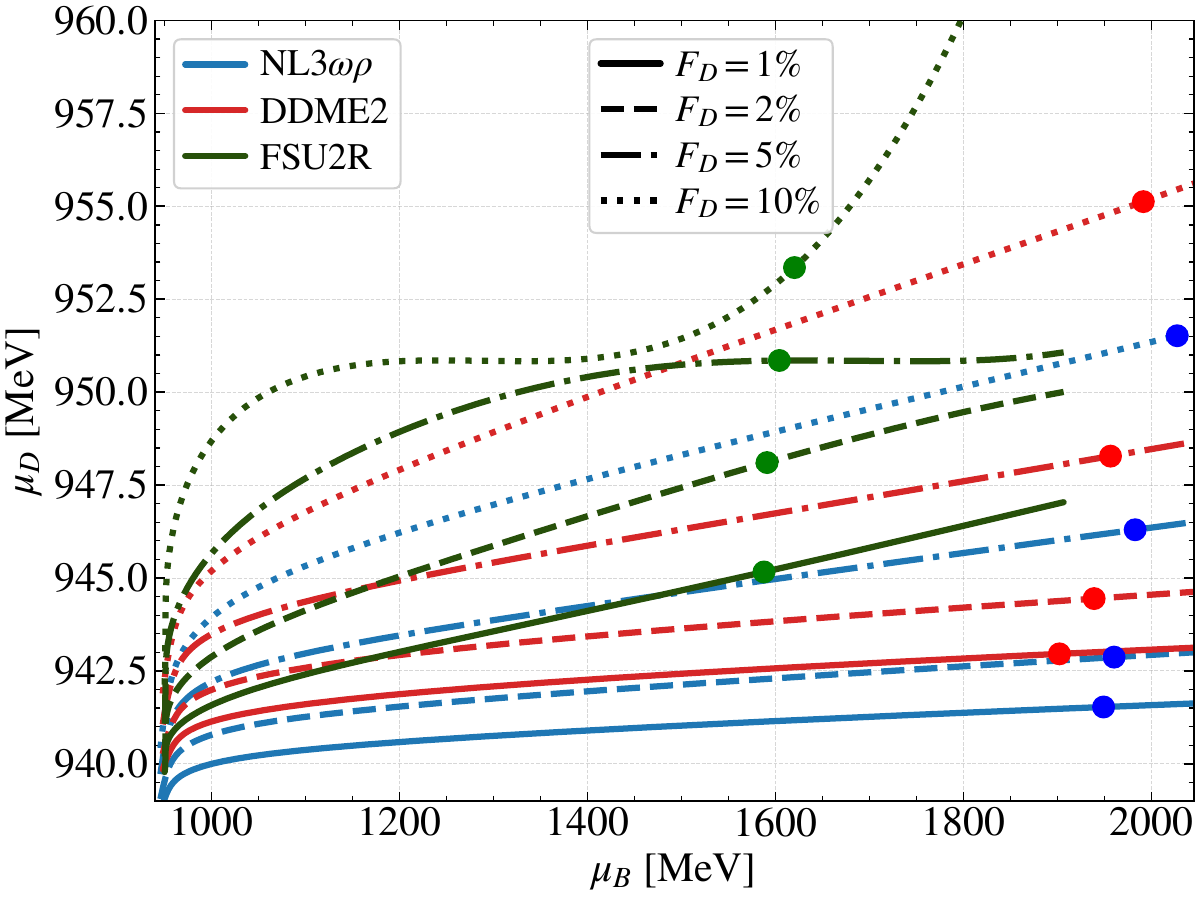}
    	\caption{Relationship between the DM and BM chemical potentials for the NL3$\omega\rho$, FSU2R, and DDME2 EOSs at different $F_D$. Colors represent the BM models, and line styles indicate the DM fractions. The evolution of $\mu_D$ with $\mu_B$ illustrates the coupling between the visible and dark sectors and its dependence on the DM content of the star. The markers on the curves denote the position of the central baryon chemical potentials of each stellar configuration along the corresponding curve.} 
\label{muBmuD_curve}
	\end{figure}
The $\mu_D$--$\mu_B$ relations in \cref{muBmuD_curve} provide a direct measurement of how the BM--DM interaction couples the thermodynamics of the two sectors. For all three EOS models, $\mu_D$ increases monotonically with $\mu_B$, reflecting the fact that both chemical potentials grow with density and are dynamically linked through the contact interaction, $\mathcal{L}_{\rm DB}$. A notable feature is that $\mu_D$ remains confined to a relatively narrow range of approximately $940$--$960$ MeV across all models and $F_D$ values, whereas $\mu_B$ spans nearly $1000$--$1800$ MeV. This behavior indicates that the dark sector remains close to its rest-mass scale throughout the stellar interior, with $\mu_D \simeq m_D$ and only moderate corrections arising from the mean fields and the BM--DM interaction. Increasing $F_D$ systematically shifts the $\mu_D$ curves upward at a given $\mu_B$, with the separation becoming more pronounced at larger $\mu_B$. This reflects the growing influence of the interaction term in the high-density regime characteristic of NS cores. The three models probe different ranges of $\mu_B$, consistent with the $\mu_c$ reported in \cref{tabmacro}: NL3$\omega\rho$ reaches the largest $\mu_B$ values, FSU2R the smallest, and DDME2 lies in between. These trends mirror the density profiles shown in \cref{nBxR}, where larger $F_D$ leads to higher $n_c$, smaller $R$, and larger $\mu_c$. The figure, therefore, provides a microscopic counterpart to the macroscopic trends observed in the stellar structure, illustrating how DM admixture drives the core toward a more compressed and chemically energetic state.

    \begin{figure}[t!]
   \centering
	\includegraphics[width=0.5\textwidth]{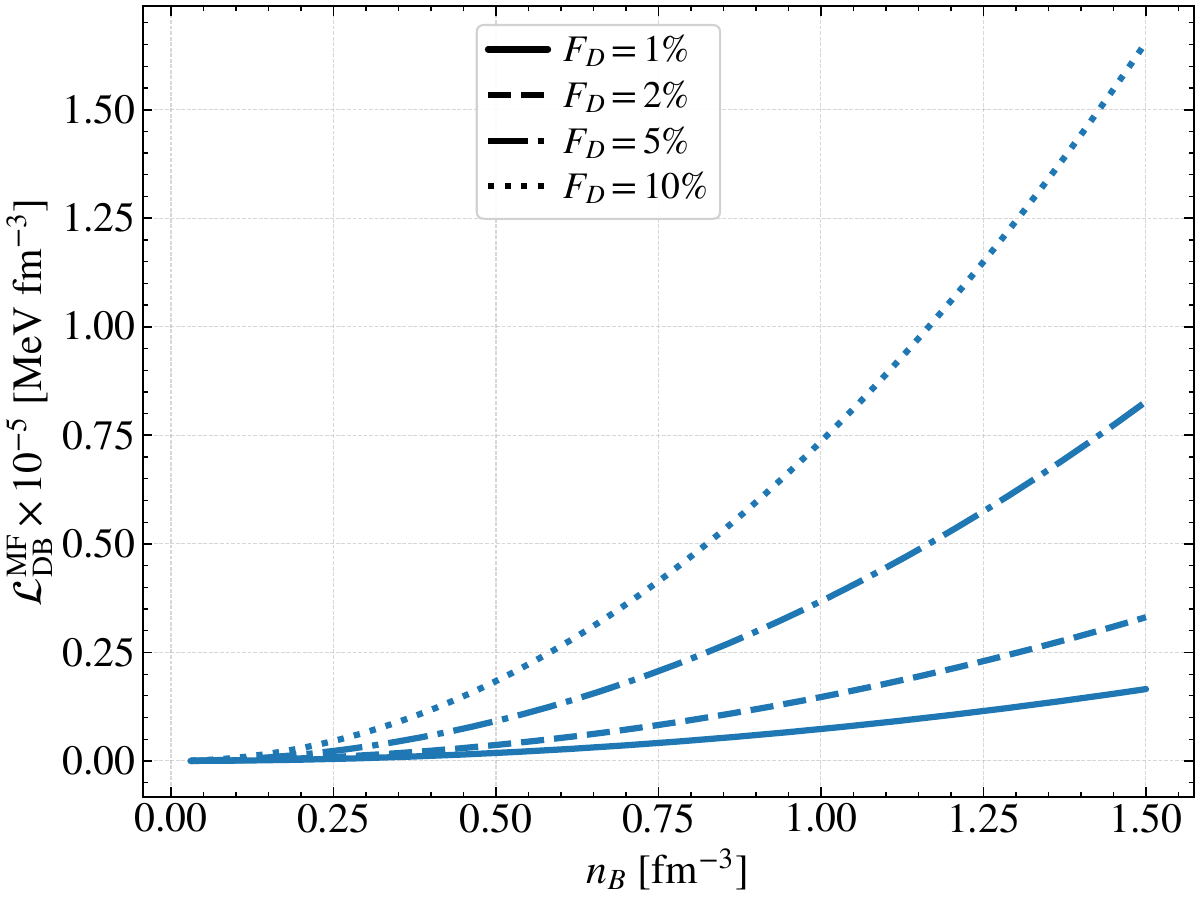}
    	\caption{The figure shows the strength of the mean-field BM--DM interaction as a function of $n_B$. The results correspond to the mean-field interaction term defined in \cref{bd}. Its functional form is independent of the EOS models.}
\label{intDB_curve}
	\end{figure}
\Cref{intDB_curve} displays the mean-field BM--DM interaction term, $\mathcal{L}_{\rm DB}^{\rm MF}=\alpha n_B n_D$, which, upon substituting $n_D=F_D n_B$, (see the discussion below \cref{NbNd}), becomes $\mathcal{L}_{\rm DB}^{\rm MF}=\alpha F_D n_B^2$. This expression shows that the BM--DM interaction is governed by the local baryon density, $\alpha$, and $F_D$ (defined in \cref{dmf}). Consequently, although the interaction strength depends explicitly on $n_B$ and $F_D$, its functional form is universal and independent of the specific EOS employed, with the EOS influencing the interaction only through the density profiles it predicts. For example, the different central densities obtained for each EOS, as shown in \cref{tabmacro}, give rise to different central BM--DM interaction strengths. Hence, the interaction grows rapidly with density, having a stronger effect in  NS cores, where both $n_B$ and $n_D$ attain their largest values. Larger values of $F_D$ further amplify this effect, producing a systematic upward shift of the interaction curves. A combined analysis based on $n_D=F_D n_B$ shows that the DM distribution follows the baryonic density profile and is concentrated toward the stellar core. The DM admixture therefore modifies the EOS and stellar structure, while the BM--DM vector interaction determines the self-consistent coupling and response of the DM within the baryonic medium. For comparison, Louren\c{c}o \textit{et al.}~\cite{Lourenco:2021dvh} considered a Higgs-portal model with an interaction energy density of order $\sim10^{-11}\,{\rm MeV\,fm^{-3}}$, whereas the present model gives $\sim10^{-6}\,{\rm MeV\,fm^{-3}}$. Although the latter is substantially larger, its contribution remains small compared with the total matter energy density. We therefore investigate the role of the interaction strength by varying $\alpha$ from zero to its experimental upper limit at fixed $F_D$, in \cref{sensitivity_test}, 
thereby distinguishing the effects of the DM abundance from those of the BM--DM coupling.}

 \begin{figure}[t!]
   \centering
	\includegraphics[width=0.5\textwidth]{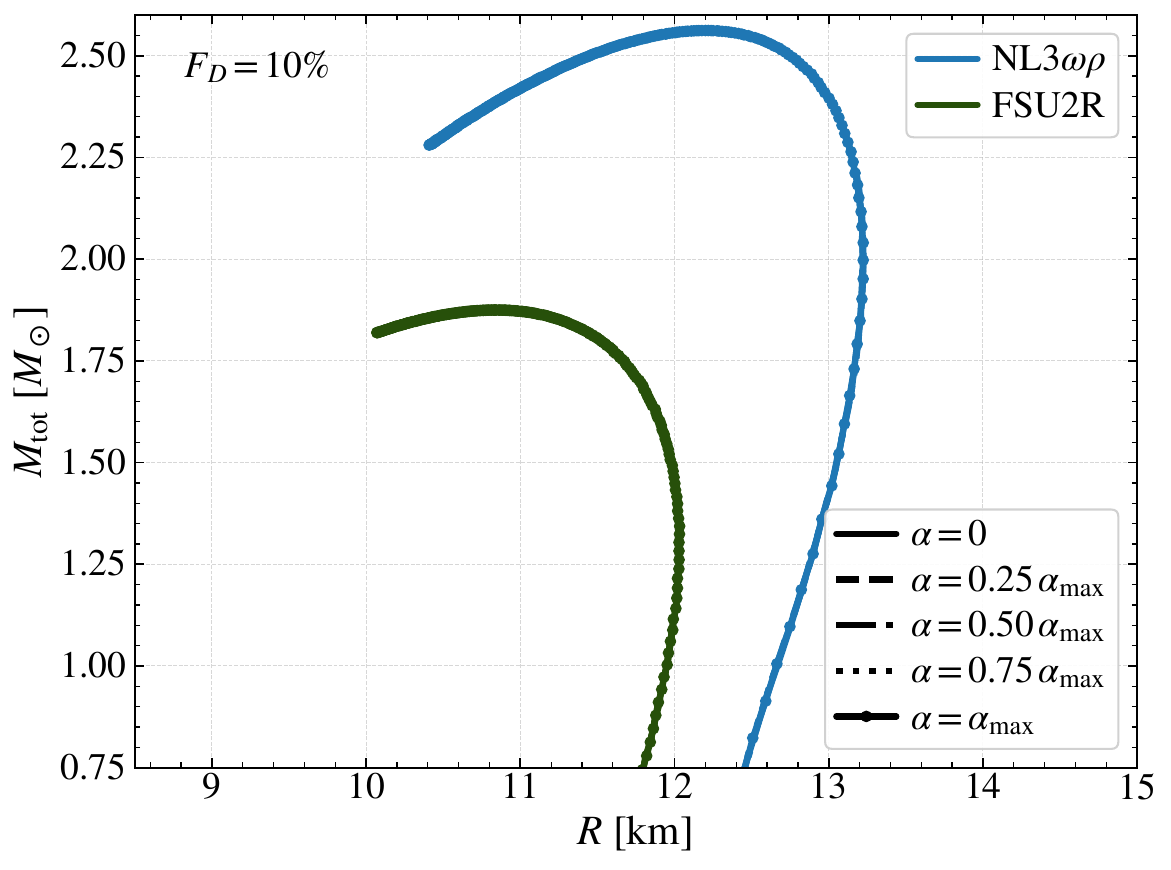}
    	\caption{Mass--radius relations for NSs with a fixed DM fraction $F_D=10\%$, shown for increasing strengths of the BM--DM vector interaction, from $\alpha=0$ to the adopted upper limit $\alpha=\alpha_{\rm max}$. The intermediate cases correspond to $\alpha=0.25\,\alpha_{\rm max}$, $0.5\,\alpha_{\rm max}$, and $0.75\,\alpha_{\rm max}$. Different colors denote the nuclear EOS models, while different line styles indicate the interaction strength relative to $\alpha_{\rm max}$. At fixed DM abundance, the overlapping curves demonstrate the weak sensitivity of the global mass--radius relation to the BM--DM interaction strength over the considered range.}
\label{sensitivity_test}
	\end{figure}
\Cref{sensitivity_test} compares the mass--radius relations for $F_D=10\%$ using the FSU2R and NL3$\omega\rho$ EOSs while varying the BM--DM vector interaction from $\alpha=0$ to the benchmark value $\alpha_{\rm max}$. The curves corresponding to different interaction strengths overlap, indicating that varying $\alpha$ within the adopted benchmark range produces no direct modification to the global stellar structure. This implies that the vector interaction governs the coupling and self-consistent response of the DM component to the baryonic medium, thereby determining how the DM is accommodated within the star, whereas the global DM fraction $F_D$ controls the amount of DM present and is consequently the dominant parameter governing the structural changes. The FSU2R EOS is shown because it exhibits the strongest sensitivity to DM admixture, while the stiffer NL3$\omega\rho$ EOS provides a comparatively less sensitive case. Thus, at fixed $F_D$, the stellar mass--radius relation is insensitive to the precise strength of the vector interaction while remaining sensitive to the underlying nuclear EOS and the amount of DM admixed.

\section{Final remarks} \label{conclusions}

In this work, we developed a self-consistent single-fluid framework for mirror DANSs that addresses two important limitations of existing approaches. Unlike models based on a fixed $k_F^\chi$, which impose an almost uniform $n_D$ and lead to a decreasing relative DM abundance toward the stellar core~\cite{Das:2020vng}, or neutron-decay-equilibrium scenarios that determine the DM population through the condition $\mu_D=\mu_n$, which may assume unrealistic maximal DM accumulation without capture dynamics (controllable by a self-interacting term ~\cite{Shirke:2023ktu}). The present framework introduces $F_D=N_D/N_B$ as a controlled global parameter. This parameter fixes the amount of DM admixed with the star while allowing the local DM density to self-consistently track $n_B$. The coupling between the two sectors is mediated by a vector current--current interaction, which generates mutual mean-field shifts in the chemical potentials and provides a direct microscopic link between the visible and dark sectors without requiring \textit{ad hoc} density prescriptions through a fixed $k_F^\chi$.

Assuming exact mirror symmetry, the dark sector possesses the same particle content, masses, and interaction structure as the BM, minimizing model dependence and allowing the coupling strength to be related directly to the spin-independent DM--nucleon scattering cross section constrained by terrestrial direct-detection experiments~\cite{CRESST:2019jnq, CRESST:2017cdd}. Thermodynamic consistency is maintained throughout the construction, including the proper treatment of rearrangement contributions in the density-dependent DDME2 model and the consistent implementation of chemical equilibrium and charge neutrality in both sectors. Applied to the NL3$\omega\rho$, NL3, FSU2R, and DDME2 EOSs {(four models of the RMF family with different symmetry energy slopes at saturation)}, the framework predicts that increasing DM content systematically softens the EOS, reduces the maximum supported mass, increases stellar compactness, suppresses the $Y_p$  at a given density, and shifts the onset of the direct Urca process to more or less massive stellar configurations, depending on the density dependence of the symmetry energy of the model. We find that $F_D$ primarily controls the modifications of the EOS and global NS structure, while the BM--DM vector interaction governs the self-consistent DM response. The framework, therefore, provides a physically motivated, thermodynamically consistent, and observationally constrained platform for exploring DM signatures in compact stars and establishing quantitative connections between NS observables and the particle physics of the dark sector.

Beyond the conceptual advantages, the framework predicts a set of robust and potentially observable consequences that emerge consistently across all EOSs considered. The presence of DM systematically reduces the binding energy and incompressibility of dense matter, leading to a softer EOS, lower $M_{\rm max}$, higher $n_c$, and more compact stellar configurations, i.e., smaller radii. Quantitatively, the maximum supported mass decreases by approximately $7\%-9\%$ as $F_D$ increases from $0\%$ to $10\%$. At fixed $F_D$, varying $\alpha$ does not show any visible effect on the stellar structure, as shown in \cref{sensitivity_test}, indicating that the direct structural contribution of the BM--DM interaction is negligible over the experimentally allowed interaction range constrained by CRESST-III. Thus, the dominant structural modifications are governed by $F_D$, while $\alpha$ controls the microscopic BM--DM coupling and the self-consistent response of the DM component within the baryonic medium as they modify the chemical potentials of both species.

At the microscopic level, the BM--DM interaction simultaneously modifies the chemical equilibrium and charge neutrality conditions, which drives matter toward a more neutron-rich composition by suppressing the proton and lepton fractions while increasing the neutron abundance, at a given density. This behavior contrasts with the trends reported in two-fluid DM studies~\cite{Issifu:2024htq, Issifu:2025jac}, where increasing DM content tends to reduce the neutron fraction and favor a larger proton abundance. The difference originates from the distinct coupling mechanisms: in the present framework, the BM--DM interaction enters directly into the chemical potentials and therefore modifies the $\beta$-equilibrium and charge neutrality conditions, whereas in two-fluid models, the two sectors interact primarily through gravity and influence the composition indirectly through changes in the stellar structure.

Equally important is the direct connection between the interaction strength and the spin-independent DM--nucleon scattering cross section constrained by terrestrial direct-detection experiments. This establishes a quantitative bridge between laboratory searches and CS physics, allowing future improvements in direct-detection sensitivities to be translated into constraints on NS structure. Conversely, precise measurements of NS masses, radii, tidal deformabilities, and thermal evolution will provide complementary probes of the dark sector. The simultaneous observation of enhanced stellar compactness, reduced maximum masses, {shifting the onset of DU cooling to higher $n_{DU}$ at a given $M_{DU}$}, and modified mass--radius relations would therefore constitute a characteristic multimessenger signature of mirror DM admixture, offering a promising avenue for testing DM physics in the extreme environment of CSs.

In summary, the present framework provides a physically motivated, thermodynamically consistent, and observationally constrained description of mirror DM in NSs. By establishing a direct connection between terrestrial DM searches and multimessenger NS observations. It offers a predictive and testable avenue for probing the dark sector in the most extreme environments known in the Universe.

\begin{acknowledgments}

A. I. acknowledges financial support from the São Paulo State Research Foundation (FAPESP), Grant Nos. 2023/09545-1 and 2025/17347-0. This work is part of the project INCT-FNA (Proc. No. 464898/2014-5) and is also supported by the National Council for Scientific and Technological Development (CNPq) under Grants No. 303490/2021-7 (D.P.M.) and 306834/2022-7 (T.F.) T. F. also thanks the financial support from the Improvement of Higher Education Personnel CAPES (Finance Code 001) and FAPESP Thematic Grants (2023/13749-1 and 2024/17816-8). F.M.S. would like to thank CNPq for financial support under research project No. 403007/2024-0 and research fellowship No. 201145/2025-1. This work was partially supported by national funds from FCT (Fundação para a Ciência e a Tecnologia, I.P, Portugal) under project UID/04564/2025, identified by DOI 10.54499/UIDB/04564/2025.
\end{acknowledgments}

\appendix

\section{Derivation of the dark matter-nucleon cross-section} 
\label{apend}
In the mirror DM scenario, the dark fermion $\psi_D$ is interpreted as a mirror nucleon with mass $m_D \simeq m_N$. Consequently, the interaction derived below corresponds to the spin-independent dark nucleon--nucleon scattering cross section constrained by direct-detection experiments. The vector--vector interaction between a dark nucleon and an ordinary nucleon is given by
\begin{equation}\label{apdx1}
    \mathcal{L}_{\rm int}=\alpha(\bar{\psi}_D\gamma^\mu\psi_D)(\bar{\psi}_N\gamma_\mu\psi_N).
\end{equation}
The elastic scattering process is
\begin{equation}
    \psi_D(p)+\psi_N(k) \rightarrow\psi_D(p')+\psi_N(k').
\end{equation}
The corresponding Feynman amplitude is
\begin{equation}
    i\mathcal{M} = i\alpha \left[\bar{u}_D(p')\gamma^\mu u_D(p)\right]\left[\bar{u}_N(k')\gamma_\mu u_N(k)\right].
\end{equation}
The spin-averaged squared amplitude is
\begin{equation}
    |\bar{\mathcal{M}}|^2=\alpha^2L_D^{\mu\nu}L_{N\,\mu\nu},
\end{equation}
where
\begin{align}
    L_D^{\mu\nu} &={\rm Tr}\left[(\slashed{p'}+m_D)\gamma^\mu(\slashed{p}+m_D)\gamma^\nu\right]\nonumber\\
    &=4\left[p'^\mu p^\nu+p^\mu p'^\nu-g^{\mu\nu}(p\!\cdot\!p'-m_D^2)\right],
\end{align}
and
\begin{align}
    L_{N\,\mu\nu} &= {\rm Tr}\left[(\slashed{k'}+m_N)\gamma_\mu(\slashed{k}+m_N)\gamma_\nu\right]
    \nonumber\\
    &=4\left[k'_\mu k_\nu+k_\mu k'_\nu-g_{\mu\nu}(k\!\cdot\!k'-m_N^2)\right].
\end{align}

Contracting the tensors yields
\begin{align}
L_D^{\mu\nu}L_{N\,\mu\nu}&=16\Big[2(p\!\cdot\!k)(p'\!\cdot\!k')+2(p\!\cdot\!k')(p'\!\cdot\!k)
\nonumber\\
&
\qquad -2m_D^2(k\!\cdot\!k')-2m_N^2(p\!\cdot\!p')+4m_D^2m_N^2\Big].
\end{align}
Defining the Mandelstam variables in the center-of-mass frame,
\begin{equation}
    s=(p+k)^2,\qquad t=(p-p')^2,\qquad u=(p-k')^2,
\end{equation}
with
\begin{equation}
    s+t+u = 2m_D^2+2m_N^2.
\end{equation}
The relevant dot products become
\begin{align}
    p\!\cdot\!k &= p'\!\cdot\!k' =\frac{s-m_D^2-m_N^2}{2}, \\ 
    p\!\cdot\!k'&=p'\!\cdot\!k=\frac{m_D^2+m_N^2-u}{2},\\
    p\!\cdot\!p'&=m_D^2-\frac{t}{2}=\frac{2m_D^2-t}{2},\\
    k\!\cdot\!k'&=m_N^2-\frac{t}{2} =\frac{2m_N^2-t}{2}.
\end{align}

Substituting these expressions, the squared amplitude becomes
\begin{align}
    |\bar{\mathcal{M}}|^2 &=8\alpha^2\Big[(s-m_D^2-m_N^2)^2+(u-m_D^2-m_N^2)^2
    \nonumber\\
    &\qquad\qquad + 2t(m_D^2+m_N^2) \Big].
\end{align}
In the non-relativistic limit relevant for the mean-field treatment,
\begin{equation}
    s \simeq (m_D+m_N)^2,\qquad u \simeq (m_D-m_N)^2, \qquad |t| \ll m^2,
\end{equation}
which yields
\begin{equation}
    |\bar{\mathcal{M}}|^2 \simeq 64\alpha^2m_D^2m_N^2.
\end{equation}
The differential cross-section is
\begin{equation}
    \frac{d\sigma}{d\Omega} =\frac{1}{64\pi^2s}\frac{|\mathbf{p}'_{\rm CM}|}{|\mathbf{p}_{\rm CM}|}|\bar{\mathcal{M}}|^2.
\end{equation}
For elastic scattering in the non-relativistic limit,
$|\mathbf{p}'_{\rm CM}|=|\mathbf{p}_{\rm CM}|$ and
$s\simeq(m_D+m_N)^2$, giving
\begin{equation}
    \sigma_{DN} = \frac{4\alpha^2m_D^2m_N^2}  {\pi(m_D+m_N)^2} = \frac{4\alpha^2\mu_{DN}^2}{\pi},
\end{equation}
where
\begin{equation}
    \mu_{DN}= \frac{m_Dm_N}{m_D+m_N}
\end{equation}
is the reduced mass of the dark nucleon--nucleon system. For the mirror DM model, where $m_D=m_N$, the coupling constant can be written as
\begin{equation}
    \alpha=\sqrt{\frac{\pi\sigma_{DN}}{m_D^2}}.
\end{equation}
This relation provides a direct connection between the effective coupling entering the stellar EOS and the dark nucleon--nucleon scattering cross section constrained by terrestrial direct-detection experiments.
\begin{table}[t!]
\centering
\caption{Current upper limits on the spin-independent dark nucleon--nucleon scattering cross section from leading direct-detection experiments.}
\begin{tabular}{|ccc|}
\hline
\bm{$m_D\,(\mathrm{GeV})$} & \textbf{Experiment} & \bm{$\sigma_{DN}\,(\mathrm{cm}^2)$} \\
\hline
$\sim 0.1$--$1$  
& CRESST-III~\cite{CRESST:2019jnq, CRESST:2017cdd} & $\sim 10^{-37}$ -- $10^{-34}$ \\
$\sim 1$--$10$     
& SuperCDMS~\cite{SuperCDMS:2018gro, SuperCDMS:2017mbc, SuperCDMS:2017nns}& $\sim 10^{-42}$ (at $\sim 8$ GeV) \\
$\sim 6$--$10$ & XENONnT~\cite{XENON:2023cxc, XENON:2025vwd} & $\sim 2.5 \times 10^{-45}$  (at $6$ GeV)\\
$\sim 30$--$50$    & LZ~\cite{LZ:2024zvo} & $2.2 \times 10^{-48}$ (at $40$ GeV) \\
$\sim 30$--$50$ & XENONnT~\cite{XENON:2023cxc, XENON:2025vwd} & $1.7 \times 10^{-47}$ (at $30$ GeV) \\
$\sim 30$--$50$ & PandaX-4T~\cite{PandaX:2024qfu} & $1.6 \times 10^{-47}$ (at $40$ GeV) \\
\hline
\end{tabular}
\label{tab:dm_constraints}
\end{table}

\section{Derivation of the DM--BM interaction pressure and energy density}\label{apdixB}

From \cref{apdx1}, we define the dark and baryonic vector currents:
\begin{equation}
j_D^\mu=\bar{\psi}_D\gamma^\mu\psi_D, \qquad j_B^\mu=\bar{\psi}_B\gamma^\mu\psi_B,
\end{equation}
such that the interaction Lagrangian density is given by
\begin{equation}
\mathcal{L}_{\rm BD}=\alpha j_D^\mu j_{B\mu}.
\end{equation}
In the mean-field approximation, only the time-like components of the currents possess non-vanishing expectation values:
\begin{equation}
\langle j_D^\mu\rangle=(n_D,0,0,0), \qquad \langle j_B^\mu\rangle=(n_B,0,0,0),
\end{equation}
where $n_D$ and $n_B$ represent the DM and BM number densities, respectively. Thus, the mean-field interaction Lagrangian simplifies to $\mathcal{L}_{\rm BD}^{\rm MF}=\alpha n_D n_B$.

The contribution of this interaction term to the total energy density $\varepsilon_{\rm tot}$ and total pressure $P_{\rm tot}$ is obtained from the canonical energy--momentum tensor:
\begin{equation}
T^{\mu\nu}=\sum_i\left[\frac{\partial \mathcal{L}}{\partial (\partial_\mu \psi_i)}\partial^\nu \psi_i+\partial^\nu \bar{\psi}_i\frac{\partial \mathcal{L}}{\partial (\partial_\mu \bar{\psi}_i)}\right]-g^{\mu\nu}\mathcal{L}.
\end{equation}
Since $\mathcal{L}_{\rm BD}$ contains no derivatives of the fermion fields, the terms in the square bracket vanish, leaving:
\begin{equation}
T^{\mu\nu}_{\rm BD}=-g^{\mu\nu}\mathcal{L}_{\rm BD}.
\end{equation}
Adopting the metric convention $g^{\mu\nu}={\rm diag}(1,-1,-1,-1)$, the interaction contribution to the energy density is given by the $T^{00}$ component:
\begin{equation}
\varepsilon_{\rm BD}=T^{00}_{\rm BD}=-g^{00}\mathcal{L}_{\rm BD}^{MF} =-\alpha n_B n_D.
\end{equation}
The corresponding contribution to the pressure follows from the spatial components:
\begin{equation}
P_{\rm BD}=\frac{1}{3}\sum_{i=1}^{3}T^{ii}_{\rm BD} = -\frac{1}{3}\sum_{i=1}^{3}g^{ii}\mathcal{L}_{\rm BD}^{MF} = +\alpha n_B n_D,
\end{equation}
where the positive sign naturally arises because $g^{ii}=-1$ for $i=1,2,3$.

\bibliography{references}

\end{document}